\documentclass[aps,prb,amsmath,amssymb,showkeys]{revtex4}
\usepackage{graphicx}% Include figure files
\usepackage{bm}% bold math
\usepackage{bbold}

\def\br{ \bm{r} }
\def\bH{ \bm{H} }
\def\bA{ \bm{A} }
\def\bM{ \bm{{\cal M}} }
\def\b0{ \bm{0} }
\def\bhx{ \hat{\bm{x}} }
\def\bhy{ \hat{\bm{y}} }
\def\bhz{ \hat{\bm{z}} }
\def\hA{ \hat{A} }
\def\bgam{ \bm{\gamma} }
\def\bGam{ \bm{\Gamma} }

\def\sgn{\, \mathrm{sgn}\, }

\def\diag{\,\mathrm{diag}\,}
\def\hbP{ \hat{\bm{P}} }
\def\hbp{ \hat{\bm{p}} }

\begin{document}
\title{Superconductivity in quantum wires: A symmetry analysis}

\author{K. V. Samokhin\footnote{e-mail: kirill.samokhin@brocku.ca}}

\affiliation{Department of Physics, Brock University, St. Catharines, Ontario L2S 3A1, Canada}
\date{\today}

\begin{abstract}
We study properties of quantim wires with spin-orbit coupling and time reversal symmetry breaking, in normal and superconducting states. Electronic band structures are classified according to quasi-one-dimensional
magnetic point groups, or magnetic classes. The latter belong to one of three distinct types, depending on the way the time reversal operation appears in the group elements. 
The superconducting gap functions are constructed using antiunitary operations and have different symmetry properties depending on the type of the magnetic point group. 
We obtain the spectrum of the Andreev boundary modes near the end of the wire in a model-independent way, using the semiclassical approach with the boundary conditions described by a phenomenological scattering matrix. 
Explicit expressions for the bulk topological invariants controlling the number of the boundary zero modes are presented in the general multiband case 
for two types of the magnetic point groups, corresponding to DIII and BDI symmetry classes.  
\end{abstract}

\keywords{noncentrosymmetric superconductors; magnetic symmetry; Andreev bound states; bulk-boundary correspondence}

\maketitle

\section{Introduction}
\label{sec: Intro}

Inspired by both the fundamental interest and also potential applications to quantum computing, the search for Majorana fermions (MFs) has become one of the central topics in 
condensed matter physics, see Refs. \onlinecite{Alicea12,LF12,ST13,Bee13} for reviews. One of the most promising routes to the MFs is based on the observation that a one-dimensional 
(1D) lattice model of a spin-polarized $p$-wave superconductor, known as the Kitaev chain,\cite{Kit01} can have unpaired, or ``dangling'', zero-energy boundary states near its ends
(although these states are not usual fermions, in particular, they have a non-Abelian exchange statistics, we will still call them MFs, following a considerable precedent in the literature). 
One can engineer a Kitaev chain-like system in a quantum wire with the spin-orbit coupling (SOC) and a sufficiently strong
time reversal (TR) symmetry breaking, in which superconductivity is induced by proximity with a conventional bulk superconductor.\cite{Lutchyn10,Oreg10} It is in this setup that experimental signatures consistent 
with the MFs have been observed, in InSb nanowires in an applied magnetic field,\cite{InSb-wire} and also in ferromagnetic chains on a superconducting Pb substrate.\cite{Fe-chain}

Both crucial ingredients of the recent MF proposals, the asymmetric SOC and TR symmetry breaking, are known to fundamentally affect superconductivity. 
The asymmetric, or Rashba, SOC (Refs. \onlinecite{Rashba-model} and \onlinecite{Manchon15}) requires the absence of inversion symmetry, which naturally occurs in a quantum wire 
placed on a substrate. It lifts the spin degeneracy of the electron states, resulting in nondegenerate Bloch bands characterized by a nontrivial momentum-space topology.
This has profound consequences for superconductivity in three-dimensional (3D) and two-dimensional (2D) materials, which have been extensively studied in the last decade, see Refs. \onlinecite{NCSC-book} and \onlinecite{Smid17} 
for reviews. On the other hand, a TR symmetry-breaking field, either intrinsic (\textit{e.g.}, the exchange field in ferromagnets) or externally applied, also lifts the spin degeneracy of the bands and 
significantly changes the symmetry properties of the Cooper pairs.\cite{SW02,FMSC-review}

The first goal of this work is to present a complete symmetry-based analysis of the electronic bands and the superconducting states in the presence of \textit{both} the asymmetric SOC and a TR symmetry-breaking field. 
Regardless of the microscopic mechanism of pairing, the symmetry approach has proven to be an extremely powerful tool in the studies of unconventional superconductivity, helping 
to identify possible stable states and determine the gap structure.\cite{Book} We focus on the quasi-1D case and develop our analysis without any model-specific assumptions, for an arbitrary number of bands. 
We emphasize, in particular, the crucial role of antiunitary symmetries in defining the proper gap function. 

Our second goal is to calculate the spectrum of the subgap Andreev bound states (ABS) near the ends of superconducting quantum wires, again in a model-independent way. The presence of these states, 
observed, \textit{e.g.}, in high-$T_c$ cuprates and other materials,\cite{ZBCP} is an important signature of an unconventional pairing. 
The ABS energies are obtained by solving the Bogoliubov-de Gennes (BdG) equations in the semiclassical, or Andreev, approximation.\cite{And64} The MFs emerge as zero-energy ABS protected by 
topology against sufficiently small perturbations. According to the bulk-boundary correspondence principle, the number of the boundary zero modes is determined by
a certain topological invariant in the bulk.\cite{Volovik-book,top-SC}  We prove this statement in the systems under consideration and present explicit expressions for the number of zero modes for different 
magnetic symmetry types.

The rest of the paper is organized as follows. In Secs. \ref{sec: magnetic classes} and \ref{sec: bands}, we introduce the 1D magnetic points groups, or magnetic classes, and develop a symmetry classification of 
quasi-1D electron band structures. In Sec. \ref{sec: superconductivity}, superconducting pairing is analyzed for different types of the magnetic classes. In Sec. \ref{sec: ABS and topology}, 
we use the semiclassical approach to calculate the ABS spectrum in a general multiband quasi-1D superconductor, in particular, to count the number of zero-energy modes protected against symmetry-preserving perturbations.
Sec. \ref{sec: summary} contains a summary of our results.
Throughout the paper we use the units in which $\hbar=1$, neglecting the difference between the quasiparticle momentum and wavevector, and denote the absolute value of the electron charge by $e$.

\section{Magnetic classes in one dimension}
\label{sec: magnetic classes}

We consider a quasi-1D wire oriented along the $x$ direction on a $xy$-plane substrate. 
The full 3D potential energy $U(x,y,z)$ affecting the electrons is periodic in $x$, with the period $d$, but confining in both $y$ and $z$ directions. This system lacks an inversion center, 
because the substrate breaks the $z\to-z$ mirror reflection symmetry. In the presence of a uniform external magnetic field $\bH=\bm{\nabla}\times\bA$, the single-particle Hamiltonian has the following form:
\begin{equation}
\label{general H}
    \hat H=\frac{\hbP^2}{2m}+U(\br)+\frac{1}{4m^2c^2}\hat{\bm{\sigma}}[\bm{\nabla}U(\br)\times\hbP]+\mu_B\hat{\bm{\sigma}}\bH.
\end{equation}
Here $\hbP=\hbp+(e/c)\bA(\br)$ and $\hbp=-i\bm{\nabla}$ are the kinetic and canonical momenta operators, respectively, $\hat{\bm{\sigma}}=(\hat\sigma_1,\hat\sigma_2,\hat\sigma_3)$ are the Pauli matrices, 
and $\mu_B$ is the Bohr magneton. The third term describes the SOC of electrons with 
the potential $U(\br)$ and the last term is the Zeeman interaction. It is convenient to choose the vector potential in the form $\bA=\bA(y,z)$ (for instance, $A_x=H_yz-H_zy$, $A_y=-H_xz$, $A_z=0$), 
then the quasimomentum along $x$ is conserved and the Bloch states are labelled by 
the wavevector $\bm{k}=k_x\bhx$, which takes values in the first Brillouin zone (BZ): $-G/2<k_x\leq G/2$. Here $G=2\pi/d$ is the basis vector of the 1D reciprocal lattice. 

The rotations and reflections leaving $U(\br)$ invariant form a point group $\mathbb{G}$. In the quasi-1D case, there are just two basic point-group operations -- the mirror reflections $\sigma_x$ and $\sigma_y$, 
which act as follows: $\sigma_xU(x,y,z)=U(-x,y,z)$ and $\sigma_yU(x,y,z)=U(x,-y,z)$, while their product is equivalent to a $\pi$-rotation about the $z$ axis: $\sigma_x\sigma_yU(x,y,z)=U(-x,-y,z)=C_{2z}U(x,y,z)$ 
(our notations for the symmetry operations are the same as in Ref. \onlinecite{LL-3}). It is easy to see that there are only five quasi-1D point groups: 
$\mathbf{C}_1=\{E\}$, $\mathbf{D}_x=\{E,\sigma_x\}$, $\mathbf{D}_y=\{E,\sigma_y\}$, $\mathbf{C}_2=\{E,\sigma_x\sigma_y\}$, and $\mathbf{V}=\{E,\sigma_x,\sigma_y,\sigma_x\sigma_y\}$, see Ref. \onlinecite{Sam17}.

The symmetry analysis can be extended to include the quasi-1D systems in which the normal state breaks TR symmetry. In this case, the crystal symmetry 
is characterized by one of the \textit{magnetic} point groups, or the magnetic classes, $\mathbb{G}_M$, whose elements leave both $U(\br)$ and $\bH$ invariant.\cite{LL-8} 
We use the notation $\bH$ in a broad sense, for any spatially uniform TR symmetry-breaking field, which transforms as a pseudovector under the reflections, \textit{i.e.}, 
$\sigma_x\bH=(H_x,-H_y,-H_z)$ and $\sigma_y\bH=(-H_x,H_y,-H_z)$, 
and also changes sign under the TR operation ${\cal K}$. In a ferromagnetic system, $\bH$ can be taken to represent the exchange field or the spontaneous magnetization.

According to general theory,\cite{LL-8} the magnetic classes can be grouped into three types. Type I classes do not involve the TR operation at all, neither by itself nor in combination 
with the reflections $\sigma_x$ and $\sigma_y$ and, therefore are formally the same as the ordinary point groups, \textit{i.e.}, $\mathbb{G}_M=\mathbb{G}$. 
These classes, which are listed in Table \ref{table: Type I}, describe TR symmetry-breaking systems with $\bH\neq\b0$, the only exception being $\mathbb{G}_M=\mathbf{V}$, which does not allow the existence 
of a uniform pseudovector $\bH$. Type II classes contain the TR operation ${\cal K}$ itself and, therefore describe TR invariant systems, with $\bH=\b0$. These classes are obtained from the ordinary point 
groups as direct products $\mathbb{G}_M=\mathbb{G}\times{\cal K}$.
Finally, Type III magnetic classes contain the TR operation only in combination with the reflections $\sigma_x$ and $\sigma_y$. In the notation for these classes, 
$\mathbb{G}_M=\mathbb{G}(\tilde{\mathbb{G}})$, the unitary subgroup $\tilde{\mathbb{G}}$ includes all elements of $\mathbb{G}$ which are not multiplied by the antiunitary operation ${\cal K}$. 
In order to obtain Type III magnetic classes, one has to identify all subgroups $\tilde{\mathbb{G}}$ of index 2 for each ordinary point group $\mathbb{G}$. Then, 
$\mathbb{G}_M=\tilde{\mathbb{G}}+({\cal K}\tilde g)\tilde{\mathbb{G}}$, where $\tilde g$ is any element of $\mathbb{G}$ that is not in $\tilde{\mathbb{G}}$.
The resulting list is shown in Table \ref{table: Type III}. 

Thus we conclude that there are altogether sixteen quasi-1D magnetic classes of all three types.  
One can see from Tables \ref{table: Type I} and \ref{table: Type III} that, if TR symmetry is broken by a uniform field $\bH$, then there are ten possible ``ferromagnetic'' classes: 
\begin{equation}
\label{FM-classes}
  \mathbb{G}_M=\mathbf{C}_1, \mathbf{D}_x, \mathbf{D}_y, \mathbf{C}_2, \mathbf{D}_x(E), \mathbf{D}_y(E), \mathbf{C}_2(E), \mathbf{V}(\mathbf{D}_x), \mathbf{V}(\mathbf{D}_y), \mathbf{V}(\mathbf{C}_2).
\end{equation}
For a given orientation of the field, only certain symmetries can be realized. In general, \textit{i.e.}, if all three components of $\bH$ are nonzero, then the magnetic class is $\mathbf{C}_1$ (Type I), 
regardless of the symmetry of the potential $U(\br)$. If the magnetic field is along a high-symmetry direction, then the number of possibilities increases. 
For example, $\bH\parallel\bhx$ can be realized in five classes: $\mathbf{C}_1$, $\mathbf{D}_x$, $\mathbf{D}_y(E)$, $\mathbf{C}_2(E)$, and $\mathbf{V}(\mathbf{D}_x)$,
while $\bH\parallel\bhz$ can be realized in $\mathbf{C}_1$, $\mathbf{C}_2$, $\mathbf{D}_x(E)$, $\mathbf{D}_y(E)$, and $\mathbf{V}(\mathbf{C}_2)$.

Diagonalization of the Hamiltonian (\ref{general H}) produces the Bloch states $|k_x,n\rangle$, labelled by the band index $n$, with the corresponding dispersions $\xi_n(k_x)$ 
(the energies are counted from the chemical potential and the difference between 
the latter and the Fermi energy $\epsilon_F$ is neglected). Due to the absence of inversion symmetry, the bands are nondegenerate almost everywhere in the BZ, see Sec. \ref{sec: bands}, 
and, other than being periodic in the reciprocal space, $\xi_n(k_x+G)=\xi_n(k_x)$, do not have generically any additional symmetries. However, for some magnetic classes there is enough symmetry to ensure that
\begin{equation}
\label{xi-even}
  \xi_n(k_x)=\xi_n(-k_x).
\end{equation}
We call such magnetic classes ``superconducting'', because symmetric bands are favorable for the formation of the Cooper pairs with zero center-of-mass momentum through a Bardeen-Cooper-Schrieffer (BCS)-like mechanism,
see Sec. \ref{sec: superconductivity}. 
This does not mean though that superconductivity is impossible if the property (\ref{xi-even}) does not hold. For instance, some kind of a nonuniform superconducting state might be stabilized in asymmetric bands. 
We do not consider such possibilities here.  

The property (\ref{xi-even}) obviously holds for all five Type II classes, \textit{i.e.}, for 
\begin{equation}
\label{classes-even-bands-II}
  \mathbb{G}_M=\mathbf{C}_1\times{\cal K},\ \mathbf{D}_x\times{\cal K},\ \mathbf{D}_y\times{\cal K},\ \mathbf{C}_2\times{\cal K},\ \mathbf{V}\times{\cal K}, 
\end{equation}
in which case the Bloch states with opposite momenta form the Kramers pairs. For Type I and Type III magnetic classes, the bands are even 
in $k_x$ only if there is a group element which transforms $k_x$ into $-k_x$, \textit{i.e.}, at least one of $\sigma_x$, ${\cal K}\sigma_y$, or $\sigma_x\sigma_y$. It is easy to see that this requirement is satisfied for 
only three of Type I classes:    
\begin{equation}
\label{classes-even-bands-I}
  \mathbb{G}_M=\mathbf{D}_x,\ \mathbf{C}_2,\ \mathbf{V}, 
\end{equation}
and three of Type III classes:
\begin{equation}
\label{classes-even-bands-III}
  \mathbb{G}_M=\mathbf{D}_y(E),\ \mathbf{V}(\mathbf{D}_x),\ \mathbf{V}(\mathbf{C}_2). 
\end{equation}
Note that all ``superconducting'' magnetic classes correspond to $H_y=0$, therefore any deviation of the magnetic field from the $xz$ plane 
produces asymmetric bands and is, therefore detrimental for superconductivity.

\begin{table}
\caption{Type I magnetic classes in quasi-1D, with the corresponding directions of a uniform TR symmetry-breaking field. The magnetic class $\mathbf{V}$ does not allow ferromagnetism 
(\textit{i.e.}, the existence of a uniform pseudovector $\bH$).}
\begin{tabular}{|c|c|}
    \hline
    $\mathbb{G}_M$ & $\bH$  \\ \hline
    $\mathbf{C}_1=\{E\}$  &  $H_x\bhx+H_y\bhy+H_z\bhz$  \\ \hline
    $\mathbf{D}_x=\{E,\sigma_x\}$ & $H_x\bhx$  \\ \hline
    $\mathbf{D}_y=\{E,\sigma_y\}$ & $H_y\bhy$  \\ \hline
    $\mathbf{C}_2=\{E,\sigma_x\sigma_y\}$ & $H_z\bhz$ \\ \hline
    $\ \mathbf{V}=\{E,\sigma_x,\sigma_y,\sigma_x\sigma_y\}\ $ & $\b0$  \\ \hline 
\end{tabular}
\label{table: Type I}
\end{table}

\begin{table}
\caption{Type III magnetic classes in quasi-1D, with the corresponding directions of a uniform TR symmetry-breaking field.}
\begin{tabular}{|c|c|}
    \hline
    $\mathbb{G}_M$ & $\bH$  \\ \hline
    $\mathbf{D}_x(E)=\{E,{\cal K}\sigma_x\}$ & $H_y\bhy+H_z\bhz$  \\ \hline
    $\mathbf{D}_y(E)=\{E,{\cal K}\sigma_y\}$ & $H_x\bhx+H_z\bhz$  \\ \hline
    $\mathbf{C}_2(E)=\{E,{\cal K}\sigma_x\sigma_y\}$ & $H_x\bhx+H_y\bhy$ \\ \hline
    $\ \mathbf{V}(\mathbf{D}_x)=\{E,\sigma_x,{\cal K}\sigma_y,{\cal K}\sigma_x\sigma_y\}\ $ &  $H_x\bhx$  \\ \hline
    $\ \mathbf{V}(\mathbf{D}_y)=\{E,{\cal K}\sigma_x,\sigma_y,{\cal K}\sigma_x\sigma_y\}\ $ &  $H_y\bhy$  \\ \hline
    $\ \mathbf{V}(\mathbf{C}_2)=\{E,{\cal K}\sigma_x,{\cal K}\sigma_y,\sigma_x\sigma_y\}\ $ &  $H_z\bhz$  \\ \hline
\end{tabular}
\label{table: Type III}
\end{table}

\section{Electronic band structure}
\label{sec: bands}

In order to get more detailed information about the band structure, in this section we derive an effective momentum-space Hamiltonian of noninteracting electrons. We adapt the general approach developed in Ref. \onlinecite{Sam09} 
to the quasi-1D case, with modifications required in the presence of magnetic field. The starting point is Eq. (1), which can be represented in the form 
$\hat H=\hat H_s+\hat H_a$, where 
\begin{eqnarray}
\label{H-symm}
  && \hat H_s=\frac{\hbp^2}{2m}+U_s(\br)+\frac{1}{4m^2c^2}\hat{\bm{\sigma}}[\bm{\nabla}U_s(\br)\times\hbp],\\
\label{H-asymm}
  && \hat H_a=U_a(\br)+\frac{1}{4m^2c^2}\hat{\bm{\sigma}}[\bm{\nabla}U_a(\br)\times\hbp]+\frac{e}{4m^2c^3}\hat{\bm{\sigma}}[\bm{\nabla}U(\br)\times\bA]+\frac{e}{2mc}(\hbp\bA+\bA\hbp)+\frac{e^2}{2mc^2}\bA^2
      +\mu_B\hat{\bm{\sigma}}\bH.
\end{eqnarray}
Here 
$$
  U_s(\br)=\frac{U(\br)+U(-\br)}{2},\quad U_a(\br)=\frac{U(\br)-U(-\br)}{2}
$$
are respectively the inversion-symmetric and antisymmetric parts of the potential $U(\br)$. Note that $\hat H_s$ is invariant with respect to both the full 3D inversion $I$ and time reversal ${\cal K}$ 
(recall that the TR operation for spin-1/2 particles is ${\cal K}=i\hat\sigma_2{\cal K}_0$, where ${\cal K}_0$ is complex conjugation), while $\hat H_a$ has neither of these symmetries. 

Diagonalizing $\hat H_s$ one obtains the Bloch bands $\epsilon_\mu(k_x)$, with the properties $\epsilon_\mu(k_x)=\epsilon_\mu(-k_x)$ and $\epsilon_\mu(k_x+G)=\epsilon_\mu(k_x)$. 
Due to the combined symmetry ${\cal K}I$, which leaves the wavevector unchanged, the bands are twofold degenerate, and one can introduce the ``pseudospin''
index $\alpha$ to distinguish two degenerate Bloch states at each $k_x$ (Ref. \onlinecite{pseudospin}). These states, denoted by $|k_x\mu\alpha\rangle$, can be chosen to transform under TR and the point group operations 
in the same way as the pure spin eigenstates, which justifies using the same notation for the pseudospin, $\alpha=\uparrow$ or $\downarrow$, as for the usual spin.

In the gauge $\bA=\bA(y,z)$, it is easy to see that $\hat H_a$, see Eq. (\ref{H-asymm}), is lattice-periodic in the $x$ direction and, therefore diagonal in $k_x$. Its matrix elements in the pseudospin basis can be written as 
\begin{equation}
\label{Hprime-matrix}
  \langle k_x\mu\alpha|\hat H_a|k_x\nu\beta\rangle=i{\cal L}_{\mu\nu}(k_x)\delta_{\alpha\beta}+\bM_{\mu\nu}(k_x)\bm{\sigma}_{\alpha\beta}.
\end{equation}
All effects of the breaking of TR and 3D inversion symmetries are contained in the matrices ${\cal L}_{\mu\nu}$ and $\bM_{\mu\nu}=({\cal M}_{1,\mu\nu},{\cal M}_{2,\mu\nu},{\cal M}_{3,\mu\nu})$.
Thus we arrive at the following second-quantized form of the noninteracting electron Hamiltonian:
\begin{equation}
\label{H-nonint}
  \hat H=\sum_{k_x,\mu\nu}\sum_{\alpha\beta}[\epsilon_\mu(k_x)\delta_{\mu\nu}\delta_{\alpha\beta}+i{\cal L}_{\mu\nu}(k_x)\delta_{\alpha\beta}
    +\bM_{\mu\nu}(k_x)\bm{\sigma}_{\alpha\beta}]\hat a^\dagger_{k_x\mu\alpha}\hat a_{k_x\nu\beta},
\end{equation}
where $\hat a^\dagger$ and $\hat a$ are the fermionic creation and annihilation operators in the pseudospin states. The bands $\xi_n(k_x)$ are obtained by diagonalizing the above 
Hamiltonian. It is easy to see that the twofold pseudospin degeneracy at given $k_x$ is lifted only if $\bM_{\mu\nu}(k_x)\neq\b0$.

The matrices ${\cal L}_{\mu\nu}(k_x)$ and $\bM_{\mu\nu}(k_x)$ must satisfy certain conditions imposed by symmetry. It follows from the general requirements of the lattice periodicity and Hermiticity
that 
\begin{eqnarray}
\label{LM-periodic}
  && {\cal L}_{\mu\nu}(k_x)={\cal L}_{\mu\nu}(k_x+G),\quad \bM_{\mu\nu}(k_x)=\bM_{\mu\nu}(k_x+G),\\
\label{LM-Hermite}
  && {\cal L}_{\mu\nu}(k_x)=-{\cal L}_{\nu\mu}^*(k_x),\quad \bM_{\mu\nu}(k_x)=\bM_{\nu\mu}^*(k_x).
\end{eqnarray}
Under time reversal, we have
\begin{equation}
\label{LM-TR}
  {\cal K}:\ {\cal L}_{\mu\nu}(k_x)\to -{\cal L}_{\mu\nu}^*(-k_x),\quad \bM_{\mu\nu}(k_x)\to -\bM_{\mu\nu}^*(-k_x), 
\end{equation}
while under the reflections,
\begin{eqnarray}
\label{LM-sigma-x}
  && \sigma_x:\ {\cal L}_{\mu\nu}(k_x)\to {\cal L}_{\mu\nu}(-k_x),\quad \bM_{\mu\nu}(k_x)\to C_{2x}\bM_{\mu\nu}(-k_x),\\
\label{LM-sigma-y}
  && \sigma_y:\ {\cal L}_{\mu\nu}(k_x)\to {\cal L}_{\mu\nu}(k_x),\quad \bM_{\mu\nu}(k_x)\to C_{2y}\bM_{\mu\nu}(k_x).
\end{eqnarray}
Here $C_{2x}$ and $C_{2y}$ are $\pi$-rotations about the $x$ and $y$ axes, respectively.

For Type II classes, we have $\bH=\b0$ and $\bA=\b0$, therefore $\hat H_a$ is odd in $\br$. Then it follows from Eq. (\ref{Hprime-matrix}) that
\begin{equation}
\label{LM-odd-k-x}
  {\cal L}_{\mu\nu}(k_x)=-{\cal L}_{\mu\nu}(-k_x),\quad  \bM_{\mu\nu}(k_x)=-\bM_{\mu\nu}(-k_x).
\end{equation}
Using Eqs. (\ref{LM-Hermite}) and (\ref{LM-TR}), we obtain that ${\cal L}_{\mu\nu}$ is a real antisymmetric matrix, while $\bM_{\mu\nu}$ is a real symmetric matrix, at each $k_x$.     
The point group symmetries can impose additional constraints. For example, if the magnetic class is $\mathbf{D}_x\times{\cal K}$, then the invariance under $\sigma_x$ dictates that 
${\cal L}_{\mu\nu}(k_x)={\cal L}_{\mu\nu}(-k_x)$ and $\bM_{\mu\nu}(k_x)=C_{2x}\bM_{\mu\nu}(-k_x)$, 
see Eq. (\ref{LM-sigma-x}). Comparing this with Eq. (\ref{LM-odd-k-x}) yields ${\cal L}_{\mu\nu}(k_x)=0$ and ${\cal M}_{1,\mu\nu}(k_x)=0$ at all $k_x$. 
We leave it to an interested reader to derive the permitted forms of the noninteracting Hamiltonian (\ref{H-nonint}) for other magnetic classes. 

\begin{figure}
\includegraphics[width=10cm]{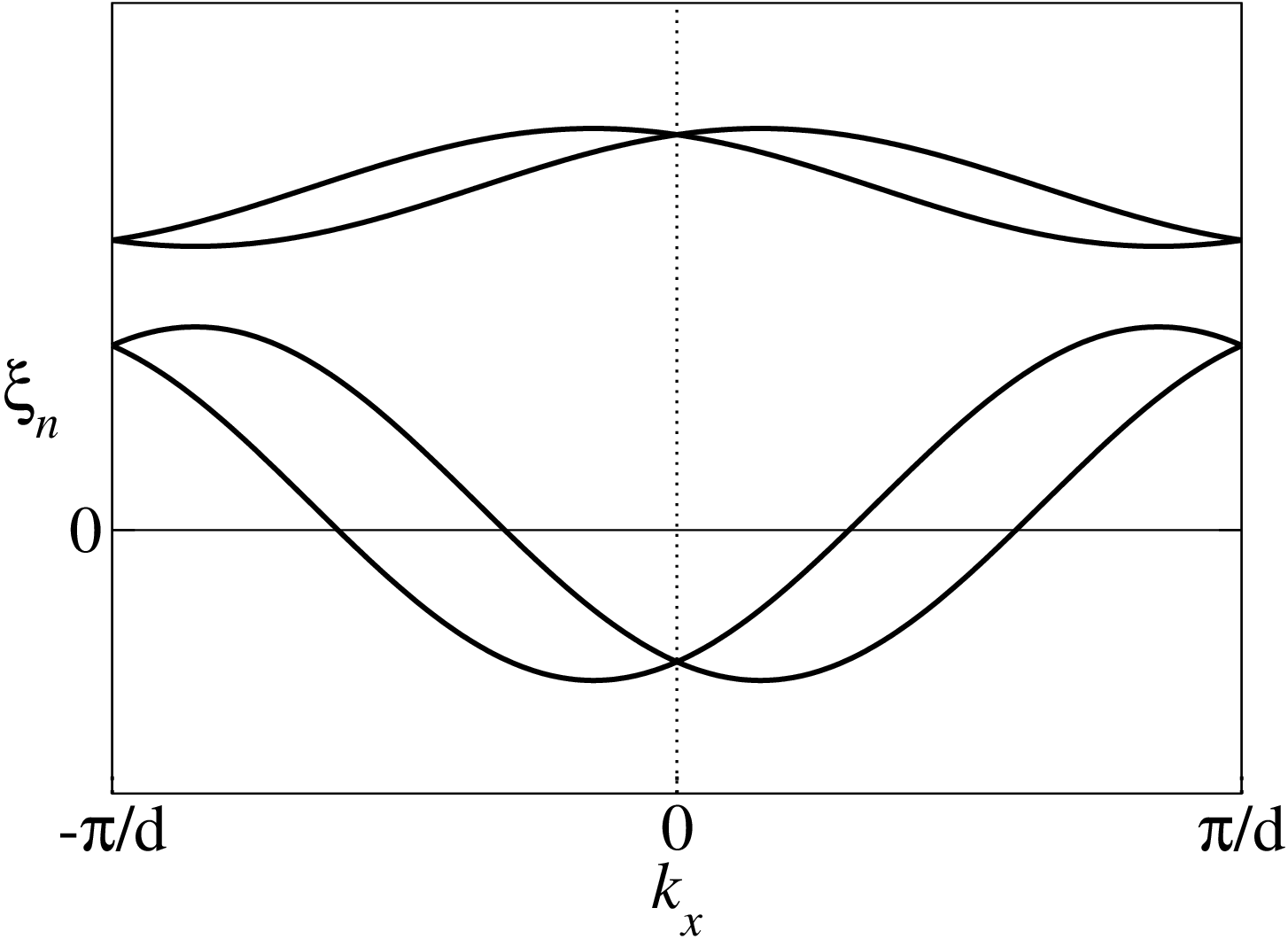}
\caption{The Bloch bands for Type II classes and the nonferromagnetic Type I class $\mathbf{V}$. The bands are symmetric and remain twofold degenerate at the TR invariant points $k_x=0$ and $\pi/d$. 
The chemical potential shown corresponds to $N=2$.}
\label{fig: bands-II}
\end{figure}

The property (\ref{LM-odd-k-x}) leads to unavoidable isolated band degeneracies. Indeed, consider the TR invariant wavevectors $k_x=K$, satisfying $-K=K+G$. There are just two such points in the 1D BZ, given by 
$K_1=0$ and $K_2=\pi/d$. At these points, we have $\bM_{\mu\nu}(K)=-\bM_{\mu\nu}(-K)=-\bM_{\mu\nu}(K+G)=-\bM_{\mu\nu}(K)$, therefore $\bM_{\mu\nu}(K_{1,2})=\b0$. For the same reason, ${\cal L}_{\mu\nu}(K_{1,2})=0$. 
Thus the bands $\xi_n(k_x)$ in a Type II system remain degenerate at $k_x=K_{1,2}$, coming in pairs connected at the center and the boundaries of the BZ, 
as shown in Fig. \ref{fig: bands-II}. The number of bands crossing the Fermi level, which we denote by $N$, is even, barring some exceptional values of the chemical potential, at which either $\xi_n(0)=0$ or $\xi_n(\pi/d)=0$.  

\begin{figure}
\includegraphics[width=10cm]{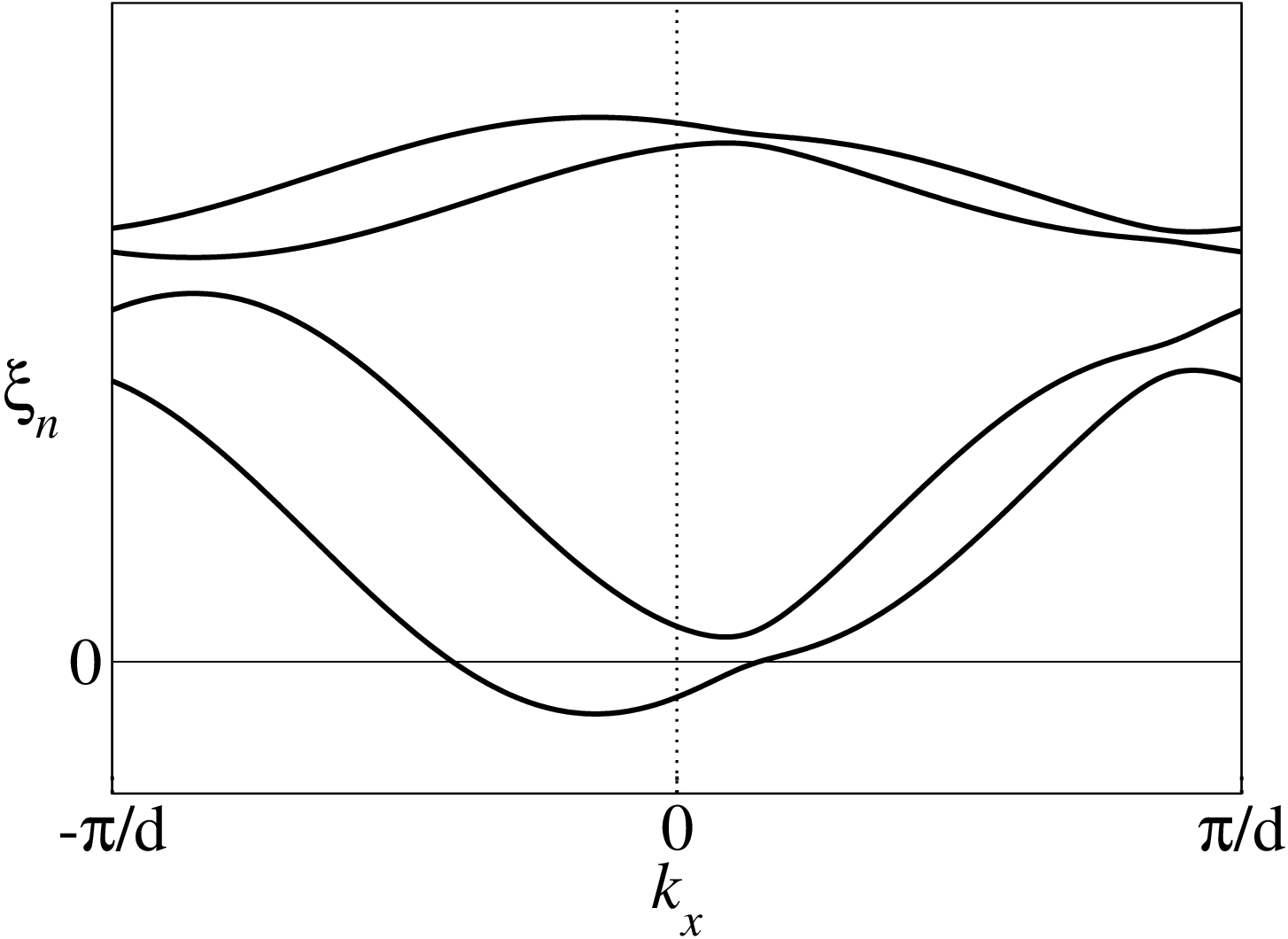}
\caption{The Bloch bands for the ``nonsuperconducting'' ferromagnetic Type I or Type III classes. 
The bands are nondegenerate everywhere in the BZ. The chemical potential shown corresponds to $N=1$.}
\label{fig: bands-I and III-asymm}
\end{figure}

\begin{figure}
\includegraphics[width=10cm]{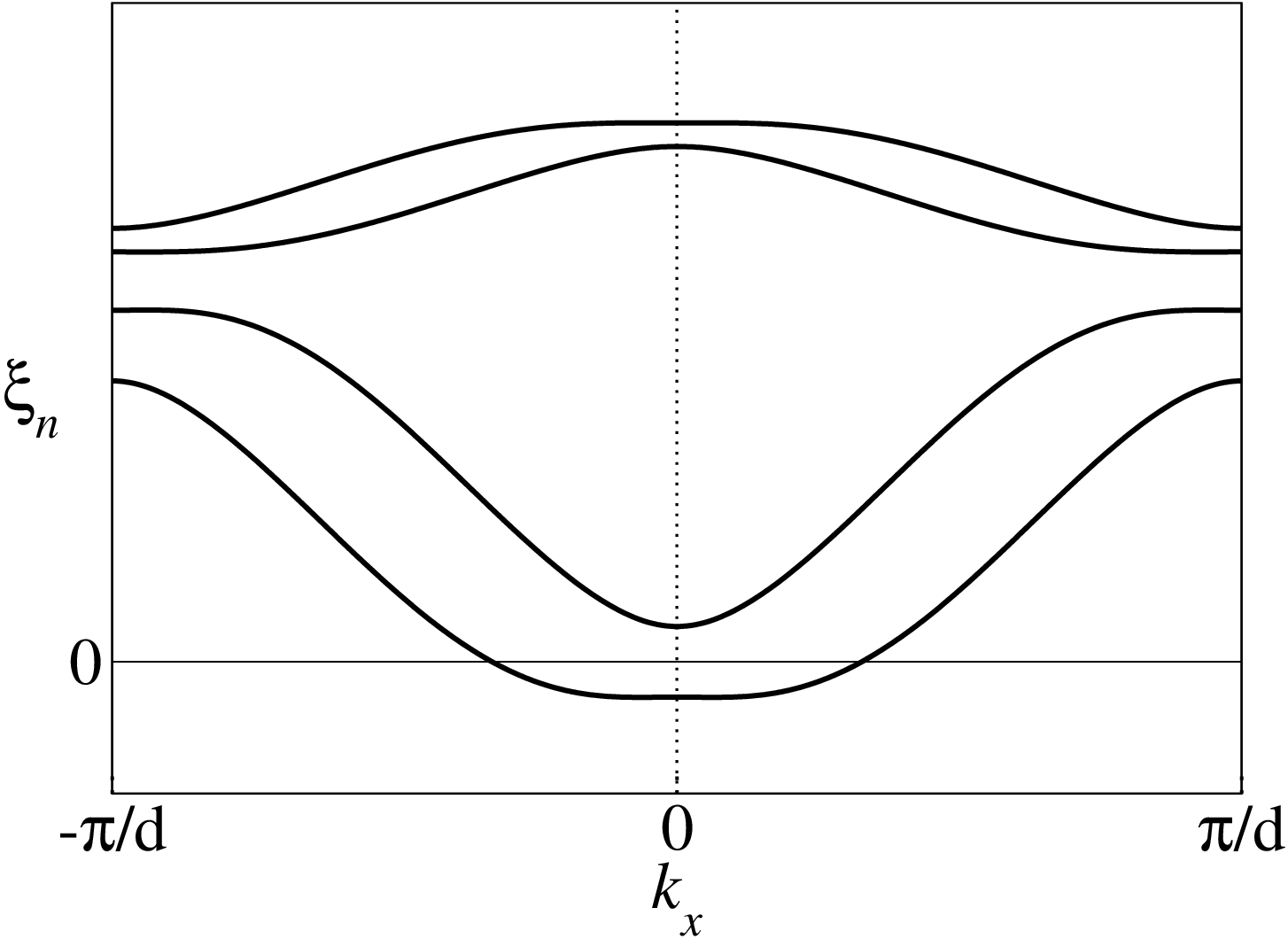}
\caption{The Bloch bands for the ``superconducting'' ferromagnetic Type I or Type III classes. 
The bands are nondegenerate everywhere in the BZ and symmetric with respect to $k_x\to-k_x$. The chemical potential shown corresponds to $N=1$.}
\label{fig: bands-I and III-symm}
\end{figure}

In contrast, for Type I and Type III magnetic classes, $\hat H_a$ is no longer odd in $\br$, and the property (\ref{LM-odd-k-x}) does not hold. 
By inspecting the magnetic point group symmetries, one can verify that there are no reasons for all elements of $\bM_{\mu\nu}$ to simultaneously vanish at $k_x=K_{1,2}$, or anywhere else in the BZ. 
The only exception is the nonferromagnetic Type I class $\mathbf{V}$, for which it follows from Eqs. (\ref{LM-sigma-x}) and (\ref{LM-sigma-y}) that ${\cal L}_{\mu\nu}(k_x)={\cal L}_{\mu\nu}(-k_x)$, 
${\cal M}_{1,\mu\nu}(k_x)=0$, ${\cal M}_{3,\mu\nu}(k_x)=0$, and ${\cal M}_{2,\mu\nu}(k_x)=-{\cal M}_{2,\mu\nu}(-k_x)$, producing the bands which are symmetric and pairwise degenerate at $k_x=K_{1,2}$, see Fig. \ref{fig: bands-II}. 
In all other Type I and Type III cases, see Eq. (\ref{FM-classes}), the band degeneracy is completely lifted at all wavevectors, and the number of bands crossing the Fermi level can be odd or even, 
as shown in Fig. \ref{fig: bands-I and III-asymm}.
For the ``superconducting'' Type I and Type III classes, listed in Eqs. (\ref{classes-even-bands-I}) and (\ref{classes-even-bands-III}), the bands are symmetric, $\xi_n(k_x)=\xi_n(-k_x)$, 
as shown in Fig. \ref{fig: bands-I and III-symm}.

\subsection{Rashba model}
\label{sec: Rashba}

The band structure peculiarities in noncentrosymmetric quasi-1D systems can be illustrated using a minimal model with just one pseudospin-degenerate band, corresponding to $\mu=0$. The general 
Hamiltonian (\ref{H-nonint}) is then reduced to
$$
  \hat H=\sum_{k_x,\alpha\beta}[\epsilon_0(k_x)\delta_{\alpha\beta}+i{\cal L}_{00}(k_x)\delta_{\alpha\beta}
    +\bM_{00}(k_x)\bm{\sigma}_{\alpha\beta}]\hat a^\dagger_{k_x\alpha}\hat a_{k_x\beta},
$$
where the band index in the fermionic operators has been dropped for brevity. It is easy to check that for all magnetic classes, $\bM_{00}$ is real, 
while ${\cal L}_{00}$ is either zero or purely imaginary, \textit{i.e.},
${\cal L}_{00}(k_x)=-i\gamma_0(k_x)$, where the even in $k_x$ part of $\gamma_0$ can be absorbed into the ``bare'' band dispersion $\epsilon_0(k_x)$. 
Introducing the notation $\bM_{00}(k_x)=\bGam(k_x)$, separating the odd and even in $k_x$ parts, and neglecting the momentum dependence of the latter, we finally arrive at the following expression:
\begin{equation}
\label{H-Rashba}
    \hat H = \sum\limits_{k_x,\alpha\beta}\left[\epsilon_0(k_x)\delta_{\alpha\beta}+\gamma_0(k_x)\delta_{\alpha\beta}+\bGam(k_x)\bm{\sigma}_{\alpha\beta}\right]\hat a^\dagger_{k_x\alpha}\hat a_{k_x\beta},
\end{equation}
where $\epsilon_0(k_x)=\epsilon_0(-k_x)$, $\gamma_0(k_x)=-\gamma_0(-k_x)$, and
\begin{equation}
\label{Gamma-def}
  \bGam(k_x)=\bgam(k_x)+\bm{h},
\end{equation}
with $\bgam(k_x)=-\bgam(-k_x)$. With the exception of the $\gamma_0$ term, the Hamiltonian (\ref{H-Rashba}) has the same form as the 1D version of the well-known Rashba model\cite{Rashba-model} in an
external magnetic field.

The first and second terms in Eq. (\ref{Gamma-def}) are usually called the ``asymmetric SOC'' and the ``Zeeman interaction'', respectively. 
The simplest form of the parameters of the Rashba model which is compatible with the reciprocal lattice periodicity is $\gamma_0(k_x)=\gamma_0\sin(k_xd)$ and $\bgam(k_x)=\bgam\sin(k_xd)$. 
The values of $\gamma_0$, $\bgam$, and $\bm{h}$ depend on the magnetic class and can be obtained by 
adapting the symmetry properties of ${\cal L}_{\mu\nu}$ and $\bM_{\mu\nu}$ discussed above, see Eqs. (\ref{LM-periodic},\ref{LM-Hermite},\ref{LM-TR},\ref{LM-sigma-x},\ref{LM-sigma-y},\ref{LM-odd-k-x}), to the one-band case. 
The results are shown in Table \ref{table: Rashba-vectors}. Note some magnetic classes are indistinguishable at the level of the Rashba model, \textit{e.g.}, $\mathbf{D}_y\times{\cal K}$, $\mathbf{V}\times{\cal K}$,
and $\mathbf{V}$. This is because the model (\ref{H-Rashba}) contains less information than the full noninteracting  Hamiltonian (\ref{H-nonint}) and, therefore cannot capture all
details of the quasi-1D electron structure.

Diagonalizing the Rashba Hamiltonian (\ref{H-Rashba}), we obtain two bands:
\begin{equation}
\label{Rashba-bands}
   \xi_\lambda(k_x)=\epsilon_0(k_x)+\gamma_0(k_x)+\lambda|\bGam(k_x)|,
\end{equation}
which are labelled by the index $\lambda=\pm$, sometimes called ``helicity'' (note that $\lambda$ is not the actual helicity, which is defined as the projection of the spin on the wavevector).
The corresponding eigenstates can be chosen in the following form:
\begin{equation}
\label{Rashba-states}
    |k_x,+\rangle=\left(\begin{array}{c}
            \cos\frac{\alpha}{2}\\ e^{i\beta}\sin\frac{\alpha}{2}
                 \end{array}\right),\quad
    |k_x,-\rangle=\left(\begin{array}{c}
            \sin\frac{\alpha}{2}\\ -e^{i\beta}\cos\frac{\alpha}{2}
                 \end{array}\right),
\end{equation}
where we used the spherical-angle parametrization: $\bGam=|\bGam|(\sin\alpha\cos\beta,\sin\alpha\sin\beta,\cos\alpha)$. 
All three types of the band structure, shown in Figs. \ref{fig: bands-II}, \ref{fig: bands-I and III-asymm}, and \ref{fig: bands-I and III-symm}, can be obtained from Eq. (\ref{Rashba-bands}) by choosing the 
appropriate values of the parameters.   
For example, according to Table \ref{table: Rashba-vectors}, in all ``superconducting'' ferromagnetic classes we have $\gamma_0=0$, $\bgam\perp\bm{h}$, and $\bm{h}\perp\bhy$. Therefore
$\xi_\lambda(k_x)=\epsilon_0(k_x)+\lambda\sqrt{\bgam^2(k_x)+\bm{h}^2}=\xi_\lambda(-k_x)$, in agreement with Eq. (\ref{xi-even}).

In the theoretical proposal of Ref. \onlinecite{Lutchyn10}, which was used in experiment to detect the MFs in semiconducting nanowires with proximity-induced superconductivity,\cite{InSb-wire} 
the external magnetic field is directed parallel to the wire, \textit{i.e.}, $\bH\parallel\bhx$. This restricts the possible magnetic symmetries to the following list: 
$\mathbb{G}_M=\mathbf{D}_x$, $\mathbf{D}_y(E)$, or $\mathbf{V}(\mathbf{D}_x)$. The actual magnetic class is determined by the details of the wire potential, the substrate, the gates, \textit{etc}. 
In Ref. \onlinecite{Lutchyn10}, the band structure was described by the Rashba model and it was assumed that the asymmetric SOC is perpendicular to the wire, namely, 
$\bgam\parallel\bhy$. According to Table \ref{table: Rashba-vectors}, this is consistent with any one of the three magnetic classes listed above.

\begin{table}
\caption{Parameters of the 1D Rashba model, see Eq. (\ref{H-Rashba}). The last column indicates whether the Rashba bands have the symmetry (\ref{xi-even}).}
\begin{tabular}{|c|c|c|c|c|}
    \hline
    $\mathbb{G}_M$  & $\gamma_0$ & $\bgam$ & $\bm{h}$ &\ SC\; \\ \hline
    $\mathbf{C}_1$  &\ $\neq 0$\ \ &   $(\gamma_1,\gamma_2,\gamma_3)$ &\ $(h_1,h_2,h_3)$\ \ & N \\ \hline
    $\mathbf{D}_x$  & $0$ & $(0,\gamma_2,\gamma_3)$ &\ $(h_1,0,0)$\ \ & Y \\ \hline
    $\mathbf{D}_y$  & $\neq 0$ & $(0,\gamma_2,0)$ &\ $(0,h_2,0)$\ \ & N \\ \hline
    $\mathbf{C}_2$  & $0$ & $(\gamma_1,\gamma_2,0)$ & $(0,0,h_3)$ & Y \\ \hline
    $\mathbf{V}$  & $0$ & $(0,\gamma_2,0)$ &\ $(0,0,0)$\ \ & Y \\ \hline
    $\ \mathbf{C}_1\times{\cal K}$\ \  & $0$ & $\ (\gamma_1,\gamma_2,\gamma_3)\ $ & $(0,0,0)$ & Y \\ \hline
    $\mathbf{D}_x\times{\cal K}$  & $0$ & $(0,\gamma_2,\gamma_3)$ & $(0,0,0)$ & Y \\ \hline
    $\mathbf{D}_y\times{\cal K}$  & $0$ & $(0,\gamma_2,0)$ &  $(0,0,0)$ & Y \\ \hline
    $\mathbf{C}_2\times{\cal K}$  & $0$ & $(\gamma_1,\gamma_2,0)$ & $(0,0,0)$ & Y \\ \hline
    $\mathbf{V}\times{\cal K}$  & $0$ & $(0,\gamma_2,0)$ &  $(0,0,0)$ & Y \\ \hline 
    $\mathbf{D}_x(E)$  & $\neq 0$ & $(0,\gamma_2,\gamma_3)$ &\ $(0,h_2,h_3)$\ \ & N \\ \hline
    $\mathbf{D}_y(E)$  & $0$ & $(0,\gamma_2,0)$ &  $(h_1,0,h_3)$ & Y \\ \hline
    $\mathbf{C}_2(E)$  & $\neq 0$ & $(\gamma_1,\gamma_2,0)$ & $(h_1,h_2,0)$ & N \\ \hline
    $\mathbf{V}(\mathbf{D}_x)$  & $0$ & $(0,\gamma_2,0)$ & $(h_1,0,0)$ & Y \\ \hline
    $\mathbf{V}(\mathbf{D}_y)$  & $\neq 0$ & $(0,\gamma_2,0)$ & $(0,h_2,0)$ & N \\ \hline
    $\mathbf{V}(\mathbf{C}_2)$  & $0$ & $(0,\gamma_2,0)$ & $(0,0,h_3)$ & Y \\ \hline
\end{tabular}
\label{table: Rashba-vectors}
\end{table}

\section{Superconductivity in 1D nondegenerate bands}
\label{sec: superconductivity}

Suppose we have a normal state described by one of the ``superconducting'' magnetic classes, see Eqs. (\ref{classes-even-bands-II}), (\ref{classes-even-bands-I}), and (\ref{classes-even-bands-III}). 
We further assume that if TR symmetry is broken, it is due to a uniform field $\bm{H}$, thus eliminating the nonferromagnetic class $\mathbf{V}$ from the consideration.  
In a BCS-like mechanism of superconductivity, the Cooper pairs are formed by quasiparticles with opposite momenta and the pairing interaction is only effective near the Fermi surface. 
The ``Fermi surface'' is given by the set of $2N$ Fermi wavevectors $\pm k_{F,n}$, which are the roots of the equations $\xi_n(k_x)=0$. Here $N$ is the number of nondegenerate bands crossing the Fermi level. 
As shown in the previous section, for Type II classes $N$ is even (Fig. \ref{fig: bands-II}), while for Type I and Type III classes $N$ can be even or odd (Fig. \ref{fig: bands-I and III-symm}).  

The exact band states $|k_x,n\rangle$, which include all effects of the lattice potential, the SOC, and the TR symmetry-breaking field, provide a natural basis for constructing the pairing interaction.
Our analysis does not rely on any specific pairing mechanism, the only assumption being that the band splitting is large enough to suppress the pairing of quasiparticles from different bands.  
The interaction Hamiltonian has the following form:
$$
    \hat H_{int}=\frac{1}{2L}\sum\limits_{k_xk_x'q}\sum_{nn'}V_{nn'}(k_x,k_x')\hat c^\dagger_{k_x+q,n}\hat c^\dagger_{-k_x,n}\hat c_{-k_x',n'}\hat c_{k_x'+q,n'},
$$
where $V_{nn}$ characterize the same-band pairing strength, $V_{nn'}$ with $n\neq n'$ describe the pair scattering between different bands, and $L$ is the length of the wire.
Treating the pairing interaction in the mean-field approximation and assuming a uniform superconducting state, we obtain:
\begin{equation}
\label{H-MF-k-k}
  \hat H=\sum_{k_x,n}\xi_n(k_x)\hat c^\dagger_{k_x,n}\hat c_{k_x,n}+\frac{1}{2}\sum_{k_x,n}\bigl[\Delta_n(k_x)\hat c^\dagger_{k_x,n}\hat c^\dagger_{-k_x,n}+\mathrm{H.c.}\bigr].
\end{equation}
The first term describes noninteracting quasiparticles, while the second term, with $\Delta_n(k_x)=-\Delta_n(-k_x)$, represents the intraband Cooper pairing 
between the states $|k_x,n\rangle$ and $|-k_x,n\rangle$. The Hamiltonian (\ref{H-MF-k-k}) can also be used to describe an ``extrinsic'' superconductivity in the wire, which is induced by proximity with 
a bulk superconductor. Unfortunately, perfunctory application of the above form of the mean-field pairing leads to a number of unpleasant consequences, which had been recognized a long time ago, see Ref. \onlinecite{Blount85}.

\subsection{General analysis}
\label{sec: BdG-general}

In order to understand the issues with the expression (\ref{H-MF-k-k}), we start with a general form of the mean-field fermionic pairing Hamiltonian in some arbitrary basis of single-particle states $|i\rangle$:
\begin{equation}
\label{H-MF-general}
  \hat H = \sum_{ij}\varepsilon_{ij}\hat c_i^\dagger\hat c_j+\frac{1}{2}\sum_{ij}\bigl(\Delta_{ij}\hat c_i^\dagger\hat c_j^\dagger+\mathrm{H.c.}\bigr)
	 = \frac{1}{2}\sum_{ij}(\hat c_i^\dagger,\ \hat c_i){\cal H}_{ij} 
	      \left(\begin{array}{c}
	      \hat c_j \\
	      \hat c_j^\dagger
	      \end{array}\right)+\mathrm{const}.
\end{equation}
Here $\varepsilon_{ij}=\langle i|\hat\varepsilon|j\rangle$ are the matrix elements of the single-particle Hamiltonian $\hat\varepsilon$ and
\begin{eqnarray}
\label{BdG-matrix-wrong}
	{\cal H}_{ij}=\left(\begin{array}{cc}
	                                            \varepsilon_{ij}  & \Delta_{ij} \\
						    \Delta^*_{ji} & -\varepsilon_{ji}
	                                            \end{array}\right).
\end{eqnarray}
Due to the anticommutation of the fermionic creation and annihilation operators, the off-diagonal terms satisfy the condition $\Delta_{ij}=-\Delta_{ji}$.

It was pointed out by Blount in Ref. \onlinecite{Blount85} that $\Delta_{ij}$ cannot 
be interpreted as a matrix element of a first-quantization operator and, therefore the matrix (\ref{BdG-matrix-wrong}) does not represent a proper BdG Hamiltonian. 
To show this, let us make a change of the single-particle basis, $|i\rangle=\sum_p|p\rangle U_{p i}$, with the coefficients $U_{p i}=\langle p|i\rangle$ 
forming a unitary matrix. The creation and annihilation operators transform as follows:
\begin{equation}
\label{ci-cmu}
  \hat c_i^\dagger=\sum_p \hat c_p^\dagger U_{p i},\quad \hat c_i=\sum_p U^*_{p i} \hat c_p,
\end{equation}
and Eq. (\ref{H-MF-general}) takes the form
\begin{eqnarray*}
  \hat H=\sum_{pq} \Bigl(\sum_{ij}U_{p i}\varepsilon_{ij}U^*_{q j}\Bigr) \hat c_p^\dagger\hat c_q
	+\frac{1}{2}\sum_{pq}\Bigl[ \Bigl(\sum_{ij}U_{p i}\Delta_{ij}U_{q j}\Bigr) \hat c_p^\dagger\hat c_q^\dagger+\mathrm{H.c.}\Bigr].
\end{eqnarray*}
The coefficients in the first term transform as expected, \textit{i.e.}, as the matrix elements of $\hat\varepsilon$:
$$
  \sum_{ij}U_{p i}\varepsilon_{ij}U^*_{q j}=\sum_{ij}\langle p|i\rangle\langle i|\hat\varepsilon|j\rangle\langle j|q\rangle=\langle p|\hat\varepsilon|q\rangle.
$$ 
However, if one tries, by analogy, to write $\Delta_{ij}=\langle i|\hat{\Delta}|j\rangle$, where $\hat\Delta$ is the ``gap operator'', then
\begin{equation}
\label{UDU}
  \sum_{ij}U_{p i}\Delta_{ij}U_{q j}=\sum_{ij}\langle p|i\rangle\langle i|\hat{\Delta}|j\rangle\langle q|j\rangle \neq \langle p|\hat{\Delta}|q\rangle.
\end{equation}
Therefore the operator $\hat{\Delta}$ does not actually exist. 

One can obtain a meaningful definition of the gap operator by modifying the pairing terms in the mean-field Hamiltonian as follows. Consider an antiunitary operation ${\cal A}$, which may or may not
be a symmetry of the system in the normal state, \textit{i.e.}, may or may not commute with $\hat\varepsilon$. We assume that ${\cal A}^2$ is either $+1$ or $-1$, when acting on spin-1/2 wave functions. 
For instance, one can use as ${\cal A}$ the time reversal operation ${\cal K}$, as in Ref. \onlinecite{Blount85}, or the latter's combination with one of the point group elements, see below. 
For each single-particle state $|i\rangle$, we introduce its ${\cal A}$-transformed counterpart ${\cal A}|i\rangle$, as well as the corresponding creation and annihilation operators  
$$
  \hat{\tilde c}_i^\dagger={\cal A}\hat c_i^\dagger{\cal A}^{-1},\qquad \hat{\tilde c}_i={\cal A}\hat c_i{\cal A}^{-1}.
$$ 
Then, for each pair of states $|i\rangle$ and $|j\rangle$ we define the gap function $\tilde\Delta_{ij}$ as a 
measure of the pairing between $|i\rangle$ and ${\cal A}|j\rangle$ and write the mean-field Hamiltonian in the form
\begin{equation}
\label{H-MF-general-corrected}
  \hat H=\sum_{ij}\varepsilon_{ij}\hat c_i^\dagger\hat c_j+\frac{1}{2}\sum_{ij}\bigl(\tilde\Delta_{ij}\hat c_i^\dagger\hat{\tilde c}_j^\dagger+\mathrm{H.c.}\bigr).
\end{equation}
It is easy to see that the gap functions $\tilde\Delta_{ij}$ have the desired transformation properties and can therefore be interpreted as matrix elements of a first-quantization operator $\hat{\tilde\Delta}$. 
Indeed, due to the antilinearity of ${\cal A}$ we have
$$
  \hat{\tilde c}_i^\dagger=\sum_p \hat{\tilde c}_p^\dagger U^*_{p i},\quad \hat{\tilde c}_i=\sum_p U_{p i} \hat{\tilde c}_p,
$$
instead of Eq. (\ref{ci-cmu}), and 
$$
  \sum_{ij}U_{p i}\tilde\Delta_{ij}U^*_{q j}=\sum_{ij}\langle p|i\rangle\langle i|\hat{\tilde\Delta}|j\rangle\langle j|q\rangle=\langle p|\hat{\tilde\Delta}|q\rangle,
$$
instead of Eq. (\ref{UDU}). 

The anticommutation of the fermionic operators imposes a certain constraint on the matrix $\tilde\Delta_{ij}$. Since the basis $|i\rangle$ is complete, one can write
\begin{equation}
\label{A-matrix}
  {\cal A}|j\rangle=\sum_i|i\rangle A_{ij},\quad A_{ij}=\langle i|{\cal A}|j\rangle.
\end{equation}
Using the antiunitary adjoint definition, $\langle i|{\cal A}|j\rangle=\langle j|{\cal A}^\dagger|i\rangle$, it is straightforward to show that the
matrix formed by the coefficients $A_{ji}$ is unitary: $\hA^{-1}=\hA^\dagger$, and also
\begin{equation}
\label{A-inverse-matrix}
  \langle i|{\cal A}^{-1}|j\rangle=\langle i|{\cal A}^\dagger|j\rangle=A_{ji}.
\end{equation}
Next, we transform the pairing terms in Eq. (\ref{H-MF-general-corrected}) as follows:
\begin{eqnarray*}
  \sum_{ij}\tilde\Delta_{ij}\hat c_i^\dagger\hat{\tilde c}_j^\dagger=\sum_{ijk}\tilde\Delta_{ij}A_{kj}\hat c_i^\dagger\hat c_k^\dagger=-\sum_{ijk}\tilde\Delta_{ij}A_{kj}\hat c_k^\dagger\hat c_i^\dagger
    =-\sum_{ijk}\tilde\Delta_{kj}A_{ij}\hat c_i^\dagger\hat c_k^\dagger.
\end{eqnarray*}
Therefore
\begin{equation}
\label{DA-AD}
  \hat{\tilde\Delta}\hA^\top=-\hA\hat{\tilde\Delta}^\top,
\end{equation}
or, in other words, the matrix $\hat{\tilde\Delta}\hA^\top=\hat{\Delta}$ is antisymmetric, which was already evident from Eq. (\ref{H-MF-general}). Depending on whether the antiunitary operation ${\cal A}$ 
squares to $+1$ or $-1$, its matrix representation is symmetric or antisymmetric. We obtain from Eq. (\ref{A-matrix}): $\langle j|{\cal A}^2|i\rangle=\langle j|\sum_{kl} A^*_{ki}A_{lk}|l\rangle=(\hA\hA^*)_{ji}$.
If ${\cal A}^2=-1$, then 
\begin{equation}
\label{A2-1-general}
  \hA^\top=-\hA,\qquad \hat{\tilde\Delta}^\top=\hA^\dagger\hat{\tilde\Delta}\hA,
\end{equation} 
but if ${\cal A}^2=+1$, then
\begin{equation}
\label{A2+1-general}
  \hA^\top=\hA,\qquad \hat{\tilde\Delta}^\top=-\hA^\dagger\hat{\tilde\Delta}\hA.
\end{equation}
Here we used the unitarity of $\hA$ and the property (\ref{DA-AD}).

One can now write the mean-field pairing Hamiltonian (\ref{H-MF-general-corrected}) in the form
\begin{equation}
\label{H-MF-tilde-rep}
  \hat H = \frac{1}{2}\sum_{ij}\left( \hat c_i^\dagger,\ \hat{\tilde c}_i \right)\tilde{\cal H}_{ij} 
	      \left(\begin{array}{c}
                  \hat c_j \\ \hat{\tilde c}_j^\dagger
                 \end{array}\right)+\mathrm{const}.
\end{equation}
The two-component fermionic operators that appear here are called the Gor'kov-Nambu operators, and the matrix connecting them is given by
\begin{equation}
\label{BdG-matrix-correct}
	\tilde{\cal H}_{ij}=\left(\begin{array}{cc}
	                                            \varepsilon_{ij}  & \tilde\Delta_{ij} \\
						    \tilde\Delta^*_{ji} & -(\hat A^\dagger\hat\varepsilon\hat A)_{ji}
	                                            \end{array}\right).
\end{equation}
In contrast to Eq. (\ref{BdG-matrix-wrong}), the last expression can be represented as $\tilde{\cal H}_{ij}=\langle i|\hat H_{BdG}|j\rangle$, \textit{i.e.},
as the matrix element of a certain first-quantization operator, which is called the BdG Hamiltonian:
\begin{equation}
\label{H-BdG}
  \hat H_{BdG}=\left(\begin{array}{cc}
                 \hat\varepsilon & \hat{\tilde\Delta} \\
		  \hat{\tilde\Delta}^\dagger & -\hat\varepsilon_{\cal A}
                 \end{array}\right).
\end{equation}
where $\hat\varepsilon_{\cal A}={\cal A}^{-1}\hat\varepsilon{\cal A}$. In particular, for the bottom-right entry we have, using Eqs. (\ref{A-matrix}), (\ref{A-inverse-matrix}), and the Hermiticity of $\hat\varepsilon$:
$$
  \langle i|\hat\varepsilon_{\cal A}|j\rangle=\langle i|{\cal A}^{-1}\sum_{kl}\varepsilon_{lk}A_{kj}|l\rangle=\sum_{kl}\varepsilon^*_{lk}A^*_{kj}A_{li}=(\hat A^\dagger\hat\varepsilon\hat A)_{ji}.
$$
The eigenstates and eigenvalues of the operator (\ref{H-BdG}) determine the wave functions and the energies of the Bogolibov quasiparticle excitations in our superconductor. 

It is easy to see that the matrices (\ref{BdG-matrix-wrong}) and (\ref{BdG-matrix-correct}) are related by a unitary transformation: ${\cal H}={\cal U}{\tilde{\cal H}}{\cal U}^{-1}$, where
$$
    {\cal U}_{ij}=\left(\begin{array}{cc}
                 \delta_{ij} & 0 \\
		  0 & A^*_{ij}
                 \end{array}\right).
$$
Therefore their eigenvalues are the same, coming in particle-hole symmetric pairs. The actual calculation has to be done using the BdG Hamiltonian (\ref{H-BdG}), whose spectrum 
can be found by utilizing various tools of quantum mechanics, \textit{e.g.}, the semiclassical approximation. Another advantage of working with the gap functions $\tilde\Delta_{ij}$ is that, unlike $\Delta_{ij}$,  
they transform in a simple way under the symmetry group operations, see the next subsection, and therefore lend themselves nicely to the standard symmetry-based analysis.\cite{Book}

\subsection{Gap symmetry in the band representation}
\label{sec: gap-band-symm}

The above arguments can be made more explicit by using the basis of the Bloch states, $|i\rangle=|k_x,n\rangle$, labelled by quasimomentum $k_x$ and the band index $n$. 
To ensure that the Cooper pairs have zero center-of-mass momentum, the antiunitary operation ${\cal A}$ should transform $k_x$ into $-k_x$, and it follows from Eq. (\ref{A-matrix}) that
\begin{equation}
\label{A-matrix-band}
  {\cal A}|k_x,n\rangle=\sum_m|-k_x,m\rangle A_{mn}(k_x),\quad A_{mn}(k_x)=\langle -k_x,m|{\cal A}|k_x,n\rangle,
\end{equation}
where the matrix $\hA(k_x)$ is unitary at each $k_x$. 

Suppose that ${\cal A}$ is a symmetry of the single-particle Hamiltonian, \textit{i.e.}, $[\hat\varepsilon,{\cal A}]=0$, therefore $\hat\varepsilon_{\cal A}=\hat\varepsilon$. 
Both $\hA(k_x)$ and the gap function matrix can be chosen in a band-diagonal form: 
\begin{equation}
\label{t_n-def}
  A_{mn}(k_x)=t_n(k_x)\delta_{mn},
\end{equation}
where $t_n$ are phase factors, and $\tilde\Delta_n(k_x)=t_n^*(k_x)\Delta_n(k_x)$. 
For a uniform superconducting state in a symmetric band, $\xi_n(k_x)=\xi_n(-k_x)$, the mean-field Hamiltonian (\ref{H-MF-general-corrected}) takes the following form:
\begin{equation}
\label{H-MF-correct}
  \hat H=\sum_{k_x,n}\xi_n(k_x)\hat c^\dagger_{k_xn}\hat c_{k_xn}+\frac{1}{2}\sum_{k_x,n}\bigl[\tilde\Delta_{n}(k_x)\hat c^\dagger_{k_xn}\hat{\tilde c}^\dagger_{k_xn}+\mathrm{H.c.}\bigr],
\end{equation}
where $\hat{\tilde c}^\dagger_{k_xn}={\cal A}\hat c^\dagger_{k_xn}{\cal A}^{-1}=t_n(k_x)\hat c^\dagger_{-k_x,n}$, and the BdG Hamiltonian (\ref{H-BdG}) can be written as
\begin{equation}
\label{H-BdG-general}
  \hat H_{BdG}(k_x)=\sum_{n} |k_x,n\rangle \left(\begin{array}{cc}
                                   \xi_n(k_x) & \tilde\Delta_n(k_x) \\
				   \tilde\Delta_n^*(k_x) & -\xi_n(k_x)
                                           \end{array}\right)
                       \langle k_x,n|.
\end{equation}
In order to make further progress, in particular, to determine the parity of the gap functions, one has to identify which ${\cal A}$ should be used for each magnetic class.

\subsubsection{Type II}
\label{sec: Type II-gap}

For Type II classes, the TR operation ${\cal K}$ is a symmetry element on its own, and it is natural to choose
\begin{equation}
\label{A-Type-II}
  {\cal A}={\cal K}.
\end{equation}
This is the standard convention in superconductivity theory, see Refs. \onlinecite{pseudospin} and \onlinecite{Blount85}. Since ${\cal A}^2={\cal K}^2=-1$, we obtain from Eq. (\ref{A2-1-general}) that $t_n(-k_x)=-t_n(k_x)$ and
\begin{equation}
\label{Delta-Type-II-parity}
  \tilde\Delta_n(k_x)=\tilde\Delta_n(-k_x),
\end{equation}
\textit{i.e.}, the gap functions in nondegenerate TR-invariant bands are even in momentum. For the Rashba model, see Eq. (\ref{Rashba-states}), the phase factors are given by $t_\lambda(k_x)=\lambda e^{-i\beta(k_x)}$. 

Now let us see how the gap functions transform under various symmetry operations. The action of an element $g$ of the point group $\mathbb{G}$ is given by
$g\hat c^\dagger_{k_xn}g^{-1}=\sum_m\hat c^\dagger_{gk_x,m}D_{mn,k_x}(g)$,
where $D$ is the unitary representation matrix of $g$ in the Bloch basis. Since $g$ is a symmetry of the single-particle Hamiltonian and $[g,{\cal K}]=0$, the representation matrix is band-diagonal and we have 
$g\hat c^\dagger_{k_xn}g^{-1}=e^{i\Phi_{k_xn}(g)}\hat c^\dagger_{gk_x,n}$ and $g\hat{\tilde c}^\dagger_{k_xn}g^{-1}=e^{-i\Phi_{k_xn}(g)}\hat{\tilde c}^\dagger_{gk_x,n}$. Under the TR operation, 
${\cal K}\hat c^\dagger_{k_xn}{\cal K}^{-1}=\hat{\tilde c}^\dagger_{k_xn}$ and ${\cal K}\hat{\tilde c}^\dagger_{k_xn}{\cal K}^{-1}=-\hat c^\dagger_{k_xn}$.
Then it follows from Eq. (\ref{H-MF-correct}) that
the gap function in each band transforms as a complex scalar, \textit{i.e.}, $\tilde\Delta_{n}(k_x)\to\tilde\Delta_{n}(g^{-1}k_x)$ under the point group symmetries and $\tilde\Delta_{n}(k_x)\to\tilde\Delta^*_{n}(k_x)$ 
under time reversal. Taken together with Eq. (\ref{Delta-Type-II-parity}), this means that the gap functions are invariant under all point group 
operations, thus corresponding to a 1D ``$s$-wave'' pairing.\cite{Sam17} Neglecting the momentum dependence of the gap functions near the Fermi points, one can write
\begin{equation}
\label{Type-II-OP}
  \tilde\Delta_{n}(k_x)=\eta_n,
\end{equation}
where the complex quantities $\eta_n$ are the components of the superconducting order parameter. Under the antiunitary symmetry operation ${\cal A}={\cal K}$ they transform as $\eta_n\to\eta^*_n$.

\subsubsection{Type III}
\label{sec: Type III-gap}
 
For all three Type III ``superconducting'' magnetic classes, see Eq. (\ref{classes-even-bands-III}), there is only one antiunitary symmetry operation which transforms $k_x$ into $-k_x$, namely,
\begin{equation}
\label{A-Type-III}
  {\cal A}={\cal K}\sigma_y.
\end{equation}
In contrast to Type II classes, now we have ${\cal A}^2=+1$. Indeed, $\sigma_y^2=(IC_{2y})^2=C_{2y}^2$, but a $2\pi$-rotation of a spin-1/2 wave function about any axis is equivalent to the multiplication by $-1$, 
therefore ${\cal K}^2\sigma_y^2=1$. Then it follows from Eq. (\ref{A2+1-general}) that $t_n(-k_x)=t_n(k_x)$ and 
\begin{equation}
\label{Delta-Type-III-parity}
  \tilde\Delta_n(k_x)=-\tilde\Delta_n(-k_x),
\end{equation}
\textit{i.e.}, the gap functions for Type III classes are odd in momentum. For the Rashba model, the action of the antiunitary opration (\ref{A-Type-III}) on the eigenstates (\ref{Rashba-states})
is equivalent to complex conjugation. By inspecting the Type III Rashba model parameters, see Table \ref{table: Rashba-vectors}, it is easy to show that $\alpha(-k_x)=\alpha(k_x)$ and $\beta(-k_x)=-\beta(k_x)$,
therefore the phase factors are given by $t_\lambda(k_x)=1$. 

While all elements of the unitary subgroup $\tilde{\mathbb{G}}$ of $\mathbb{G}_M$ act on the gap functions in the same way as in the Type II case, \textit{i.e.}, $\tilde\Delta_{n}(k_x)\to\tilde\Delta_{n}(g^{-1}k_x)$ 
under $g\in\tilde{\mathbb{G}}$, the antiunitary element acts differently: from Eq. (\ref{H-MF-correct}) we have $\tilde\Delta_{n}(k_x)\to-\tilde\Delta^*_{n}(k_x)$ under ${\cal A}={\cal K}\sigma_y$. 
According to general theory, the superconducting order parameter components can be introduced by expanding the gap functions in terms of ``building blocks'' -- 
the basis functions of the irreducible representations (IREPs). However, instead of the usual IREPs, in the present case one should use the corepresentations of the magnetic point group $\mathbb{G}_M$, 
which are derived from the IREPs of the unitary component $\tilde{\mathbb{G}}$ (Ref. \onlinecite{magn-groups}). 

It is straightforward to show that, since all group elements commute and the antiunitary elements all square to $+1$, the corepresentations of the magnetic classes (\ref{classes-even-bands-III}) are one-dimensional. 
Using the property (\ref{Delta-Type-III-parity}), we obtain:
\begin{equation}
\label{Type-III-OP}
  \tilde\Delta_{n}(k_x)=i\eta_n\phi_n(k_x),\quad \phi_n(k_x)=-\phi_n(-k_x),
\end{equation}
where $\eta_n$ are the order parameter components and $\phi_n(k_x)$ are real basis functions. The latter can be different in different bands and taken to be $\phi_n(k_x)\propto k_x$, which 
corresponds to a 1D ``$p$-wave'' pairing. Therefore a Type III superconductor can be regarded as a multiband continuum generalization of the Kitaev chain.\cite{Kit01} 
The imaginary factor in Eq. (\ref{Type-III-OP}) is introduced for convenience, to make sure that, as in the Type II case, the action of the antiunitary operation on the 
order parameter components is equivalent to complex conjugation, \textit{i.e.}, we have $\eta_n\to\eta^*_n$ under ${\cal A}={\cal K}\sigma_y$.

\subsubsection{Type I}
\label{sec: Type I-gap}

There are only two Type I magnetic classes that are simultaneously ``ferromagnetic'' and ``superconducting'', $\mathbf{D}_x$ and $\mathbf{C}_2$, see Eqs. (\ref{FM-classes}) and (\ref{classes-even-bands-I}). 
Since Type I classes do not have 
any antiunitary elements, there is no obvious choice for ${\cal A}$. While the pairing can be defined using, for instance, the TR operation, the matrix representation of ${\cal A}={\cal K}$ and the gap function 
are no longer band-diagonal, nor does the gap functions $\hat{\tilde\Delta}$ have a definite parity, see Eq. (\ref{A2-1-general}). Due to these complications, 
Type I systems have to be treated differently and will be studied elsewhere.   

We would like to note that, while it is the band representation of pairing that is best suited for the semiclassical analysis, see Sec. \ref{sec: ABS and topology}, it is possible to translate 
the results of this section into the more traditional spin representation. This can be done though only if an explicit model of the band structure is available. In Appendix \ref{app: gap spin}, 
we discuss the relation between the band and spin representations for a two-band Rashba wire.

\subsection{Stable states}
\label{sec: stable SC states}

We have shown in Sec. \ref{sec: gap-band-symm} that the superconducting order parameter in both Type II and Type III quantum wires is given by a set of $\eta_1,...,\eta_N$, which transform into their complex conjugates under the 
action of the antiunitary symmetry operation ${\cal A}$. The actual values of the order parameter components can be obtained by minimizing the Ginzburg-Landau (GL) free energy.  
Assuming a uniform state, the most general second- and fourth-order terms in the free energy density have the following form:
\begin{equation}
\label{F_GL}
  F=\sum_{mn}\alpha_{mn}\eta_m^*\eta_n+\sum_{klmn}\beta_{klmn}\eta_k^*\eta_l^*\eta_m\eta_n.
\end{equation}
In a phenomenological theory, the values of the coefficients are only constrained by the requirement that $F$ is a real scalar, which is invariant under all symmetry operations, including ${\cal A}$. 
It is then straightforward to show that the coefficients are real and satisfy $\alpha_{mn}=\alpha_{nm}$, $\beta_{klmn}=\beta_{lkmn}=\beta_{klnm}=\beta_{mnkl}$.

In the case of $N=1$, which is possible only in a Type III superconductor, one can make the single component of the order parameter real by a phase rotation, therefore 
the superconducting state is always invariant under the antiunitary operation (\ref{A-Type-III}). For $N>1$, in both Type II and Type III cases, Eq. (\ref{F_GL}) has the 
same form as the usual multiband generalization of the GL energy. The latter can have real minima, as well as intrinsically complex ones with the interband phase differences other than $0$ or $\pi$ (Ref. \onlinecite{TRSB-states}). 
Therefore in addition to the ${\cal A}$-invariant states characterized by real $\eta_n$, various antiunitary symmetry-breaking superconducting states are also phenomenologically possible.

\section{Andreev boundary modes}
\label{sec: ABS and topology}

The Bogoliubov quasiparticle energies in an infinite uniform superconductor are obtained by diagonalizing the BdG Hamiltonian (\ref{H-BdG-general}), with the following result: 
$E_n(k_x)=\pm\sqrt{\xi_n^2(k_x)+|\tilde\Delta_n(k_x)|^2}$. The gap functions are given by Eq. (\ref{Type-II-OP}) or Eq. (\ref{Type-III-OP}). While the bulk excitations are gapped, there might exist subgap states 
localized near various inhomogeneities, in particular, near the boundaries of the system. In this section, we calculate the spectrum of these states in a half-infinite ($x\geq 0$) clean wire using the semiclassical, 
or Andreev, approach.\cite{And64} To make analytical progress, we neglect self-consistency and assume that the order parameters $\eta_1,...,\eta_N$ do not depend on $x$.

\subsection{Semiclassical analysis of the boundary modes}
\label{sec: ABS}
 
The quasiparticle wave function in the $n$th band is an electron-hole spinor, which can be represented in the semiclassical approximation
as $e^{irk_{F,n}x}\psi_{n,r}(x)$, where $r=\pm$ characterizes the direction of the Fermi wavevector $rk_{F,n}$. 
The ``envelope'' function $\psi_{n,r}(x)$ varies slowly on the scale of the Fermi wavelength $k_{F,n}^{-1}$ and satisfies the Andreev equation:    
\begin{equation}
\label{And-eq-gen}
	\left(\begin{array}{cc}
		-iv_{n,r}\dfrac{d}{dx} & \tilde\Delta_n(rk_{F,n}) \\
		\tilde\Delta_n^*(rk_{F,n}) & iv_{n,r}\dfrac{d}{dx}
	\end{array}\right)\psi_{n,r}=E\psi_{n,r}.
\end{equation}
Here $v_{n,r}=(\partial\xi_n/\partial k_x)|_{k_x=rk_{F,n}}$ is the quasiparticle group velocity near the Fermi point $rk_{F,n}$ (note that $|v_{n,\pm}|=v_{F,n}$). 
Focusing on the subgap states with $|E|<|\tilde\Delta_n|$, the solution of Eq. (\ref{And-eq-gen}) has the form $\psi_{n,r}(x)=\phi(rk_{F,n})e^{-\Omega_nx/v_{F,n}}$, where 
\begin{equation}
\label{Andreev amplitude}
	\phi(rk_{F,n})=C(rk_{F,n})\left(\begin{array}{c}
		\dfrac{\tilde\Delta_n(rk_{F,n})}{E-i\Omega_n\sgn v_{n,r}} \\ 1
	\end{array}\right),
\end{equation}
$\Omega_n=\sqrt{|\tilde\Delta_n(rk_{F,n})|^2-E^2}$, and $C$ is a coefficient.

The semiclassical approximation breaks down near the boundary due to a rapid variation of the confining potential, which causes elastic transitions between the states corresponding to different Fermi wavevectors.
This can be described by an effective boundary condition for the Andreev wave functions, which is obtained as follows. Depending on the sign of the group velocity, 
the Fermi wavevectors are classified as either incident, for which $v_{n,r}<0$, or reflected, for which $v_{n,r}>0$. 
We denote the former $k^{\mathrm{in}}_{1},...,k^{\mathrm{in}}_N$ and the latter $k^{\mathrm{out}}_{1},...,k^{\mathrm{out}}_N$, with $k^{\mathrm{out}}_{n}=-k^{\mathrm{in}}_{n}$.
From Eq. (\ref{Andreev amplitude}), the Andreev wave functions at $x=0$ corresponding to the incident and reflected wavevectors can be written as
\begin{equation}
\label{phi-in-out}
  \phi(k^{\mathrm{in}}_n)=C(k^{\mathrm{in}}_n)\left(\begin{array}{c}
		\alpha^{\mathrm{in}}_n \\ 1
	\end{array}\right),\quad
  \phi(k^{\mathrm{out}}_n)=C(k^{\mathrm{out}}_n)\left(\begin{array}{c}
		\alpha^{\mathrm{out}}_n \\ 1
	\end{array}\right),
\end{equation}
where
$$
  \alpha^{\mathrm{in}}_n=\frac{\tilde\Delta_n(k^{\mathrm{in}}_{n})}{E+i\sqrt{|\tilde\Delta_n(k^{\mathrm{in}}_{n})|^2-E^2}},\quad 
  \alpha^{\mathrm{out}}_n=\frac{\tilde\Delta_n(k^{\mathrm{out}}_{n})}{E-i\sqrt{|\tilde\Delta_n(k^{\mathrm{out}}_{n})|^2-E^2}}.
$$
According to Ref. \onlinecite{Shel-bc}, the boundary conditions have the form of a linear relation between the ``in'' and ``out'' Andreev amplitudes:
\begin{equation}
\label{Shelankov-bc}
  \phi(k^{\mathrm{out}}_n)=\sum_{m=1}^N S_{nm}\phi(k^{\mathrm{in}}_{m}),
\end{equation}
with the coefficients $S_{nm}$ forming a unitary $N\times N$ matrix. The $S$ matrix is an electron-hole scalar, determined by the details of the boundary scattering at the Fermi level in the normal state. 
If a manageable microscopic description of the electron band structure is available, then the $S$ matrix can be calculated analytically, see Ref. \onlinecite{S-matrix-Rashba}. In general, it should be regarded 
as a phenomenological input. 

Inserting the expressions (\ref{phi-in-out}) into the boundary conditions (\ref{Shelankov-bc}), we obtain a system of $2N$ linear equations for $C(k^{\mathrm{in}}_n)$ and $C(k^{\mathrm{out}}_n)$. 
It has a nontrivial solution if
\begin{equation}
\label{ABS-energy-equation}
   \det\hat W(E)=0,
\end{equation}
where
$$
  \hat W(E)=\hat S-\hat M_{\mathrm{out}}^\dagger(E)\hat S\hat M_{\mathrm{in}}(E),
$$
with $\hat M_{\mathrm{in}}=\diag(\alpha^{\mathrm{in}}_1,...,\alpha^{\mathrm{in}}_N)$ and $\hat M_{\mathrm{out}}=\diag(\alpha^{\mathrm{out}}_1,...,\alpha^{\mathrm{out}}_N)$. Below we focus on calculating the 
number of the zero-energy solutions of Eq. (\ref{ABS-energy-equation}) in an ${\cal A}$-invariant superconducting state.

\subsection{Counting the zero modes}
\label{sec: zero modes}

The number of the ABS zero modes localized near $x=0$ is given by
\begin{equation}
\label{number-zero-modes}
    {\cal N}_0=\dim\ker\hat W(0).
\end{equation} 
Using Eqs. (\ref{Delta-Type-II-parity}) and (\ref{Delta-Type-III-parity}), we find
$\hat W(0)=\hat S\pm \hat{\tilde P}^\dagger\hat S\hat{\tilde P}$, where the upper (lower) sign corresponds to superconductors with a Type II (Type III) symmetry and
$$
  \hat{\tilde P}=\diag\left[\frac{\tilde\Delta_1(k^{\mathrm{out}}_{1})}{|\tilde\Delta_1(k^{\mathrm{out}}_{1})|},...,\frac{\tilde\Delta_N(k^{\mathrm{out}}_{N})}{|\tilde\Delta_N(k^{\mathrm{out}}_{N})|}\right].
$$
In terms of the order parameter components, as defined by Eqs. (\ref{Type-II-OP}) and (\ref{Type-III-OP}), we obtain:
\begin{equation}
\label{W-Type-II-III}
  \hat W(0)=\hat S\pm \hat P^\dagger\hat S\hat P,
\end{equation}
where 
$$
  \hat P=\diag\left(\frac{\eta_1}{|\eta_1|},...,\frac{\eta_N}{|\eta_N|}\right).
$$
In the Type III case, we neglected the momentum dependence of the ``$p$-wave'' basis functions near the Fermi points and used the convention $\phi_n(k_x)=\sgn(k^{\mathrm{out}}_{n})\sgn(k_x)$ in Eq. (\ref{Type-III-OP}).
Thus the sign of $\eta_n$ is defined by the value of the gap function $\tilde\Delta_n$ at the reflected wavevector $k_x=k^{\mathrm{out}}_{n}$. 

We consider only the superconducting states in which the antiunitary symmetry ${\cal A}$ is not broken. For Type II classes, these are the usual TR invariant states, in which the order parameter 
components $\eta_n$ can be made real by a gauge transformation. For Type III classes, the states invariant under ${\cal A}={\cal K}\sigma_y$ are also characterized by a real order parameter.  
We label the bands in such a way that the first $N_+$ bands have positive $\eta_n$, while the remaining $N_-=N-N_+$ bands have negative $\eta_n$, so that $\hat P=\diag(\bm{1}_+,-\bm{1}_-)$. Here and below, vectors 
subscripted with ``$+$'' (``$-$'') have $N_+$ ($N_-$) components.
Representing the $S$ matrix in the block form as
\begin{equation}
\label{S-R}
  \hat S=\left(\begin{array}{cc}
          \hat R_{++} & \hat R_{+-} \\
          \hat R_{-+} & \hat R_{--}  \\ 
         \end{array}\right),
\end{equation}
where $\hat R_{ss'}$ is a $N_{s}\times N_{s'}$ matrix ($s,s'=\pm$), Eq. (\ref{W-Type-II-III}) takes the following form:
\begin{equation}
\label{W-Type-II-R}
  \hat W(0)=2\left(\begin{array}{cc}
          \hat R_{++} & 0 \\
          0 & \hat R_{--}  \\ 
         \end{array}\right)\qquad (\mathrm{Type\ II}),
\end{equation}
or
\begin{equation}
\label{W-Type-III-R}
  \hat W(0)=2\left(\begin{array}{cc}
          0 & \hat R_{+-} \\
          \hat R_{-+} & 0 \\ 
         \end{array}\right)\qquad (\mathrm{Type\ III}).
\end{equation}
Further steps require a more detailed knowledge of the $S$ matrix. The latter is a normal-state property and has to satisfy, in addition to unitarity, certain constraints imposed by the 
antiunitary symmetry.

\subsubsection{Type II}
\label{sec: ZEABS Type II}

As shown in Sec. \ref{sec: bands}, in Type II systems $N$ is even and, therefore $N_+$ and $N_-$ have the same parity. We obtain from Eqs. (\ref{number-zero-modes}) and (\ref{W-Type-II-R}) that 
$$
  {\cal N}_0=\dim\ker\hat R_{++}+\dim\ker\hat R_{--}.
$$
According to Appendix \ref{app: S-matrix-symmetry}, in the Type II case the scattering matrix can be made antisymmetric, therefore both $\hat R_{++}$ and $\hat R_{--}$ are antisymmetric. If $N_+$ and $N_-$ are even, 
then there is no reason for $\hat R_{++}$ and $\hat R_{--}$ to have zero eigenvalues. However, if $N_+$ and $N_-$ are odd, then both $\hat R_{++}$ and $\hat R_{--}$ are singular, \textit{i.e.}, 
have at least one zero eigenvalue each. Thus we arrive at the following result:
\begin{equation}
\label{N0-Type II}
  {\cal N}_0=\left\{\begin{array}{ll}
                    2,\quad & \mathrm{if}\ N_+,N_-=\mathrm{odd},\\
		    0,\quad & \mathrm{if}\ N_+,N_-=\mathrm{even}, 
                    \end{array}\right.
\end{equation}
which means that the zero-energy ABS can only exist if there is an odd number of bands with the same sign of the gap. These zero modes form one Kramers pair per each end of the wire.

The bulk-boundary correspondence principle states that the number of the boundary zero modes is related 
to a topological invariant in the bulk.\cite{Volovik-book,top-SC} Since the antiunitary symmetry in Type II case is the usual time reversal, which squares to $-1$, our system belongs to the symmetry class DIII
and can be characterized in 1D by a $\mathbb{Z}_2$ invariant.\cite{tenfold-way} According to Ref. \onlinecite{QHZ10}, this invariant is given by
$\prod_{n=1}^N\sgn(\eta_n)=(-1)^{N_-}=(-1)^{N_+}$, therefore the states with $N_+$ and $N_-$ odd are $\mathbb{Z}_2$-nontrivial and should have a pair of the ABS zero modes, in agreement with Eq. (\ref{N0-Type II}).
If TR symmetry is broken, either due to a magnetic boundary scattering or intrinsically in the superconducting state, then the Kramers pairs of the zero modes are split. The same effect is produced by TR 
symmetry-breaking fluctuations, even if the mean-field state is TR invariant (Ref. \onlinecite{ST17}).

\subsubsection{Type III}
\label{sec: ZEABS Type III}

In Type III systems, there can be any number of bands crossing the Fermi level. As shown in Appendix \ref{app: S-matrix-symmetry}, the scattering matrix is now symmetric, therefore $\hat R_{-+}=\hat R_{+-}^\top$. 
The number of the zero modes of Eq. (\ref{W-Type-III-R}) can be calculated using the rank-nullity theorem. Namely, we write the zero-mode 
eigenvectors of $\hat W(0)$ in the form $(\bm{v}^\top_+,\bm{v}^\top_-)$, where $\bm{v}_+$ and $\bm{v}_-$ satisfy the equations 
\begin{equation}
\label{zero-modes-eqs-Type-III}
  \hat R_{+-}\bm{v}_-=\bm{0}_+,\quad \hat R_{-+}\bm{v}_+=\bm{0}_-.
\end{equation}
If $N_+>N_-$, then $\bm{v}_-=\bm{0}_-$ and the second equation (\ref{zero-modes-eqs-Type-III}) has $N_+-N_-$ nonzero solutions. 
If $N_+<N_-$, then $\bm{v}_+=\bm{0}_+$ and the first equation (\ref{zero-modes-eqs-Type-III}) has $N_--N_+$ nonzero solutions. Finally, if $N_+=N_-$, then the equations (\ref{zero-modes-eqs-Type-III}) generically have no 
nontrivial solutions. Collecting everything together, we obtain:
\begin{equation}
\label{N0-Type III}
  {\cal N}_0=|N_+-N_-|,
\end{equation}
\textit{i.e.}, there can be any integer number of the zero-energy ABS. 

The result (\ref{N0-Type III}) can also be understood using the MF language. Consider $N$ Kitaev chains in the nontrivial phase,\cite{Kit01} which is the lattice version of our quasi-1D $N$-band Type III superconductor.
The chains are labelled by the index $n$, while the lattice sites are labelled by the discrete position $1\leq x\leq L$ (we assume a half-infinite geometry, with $L\to\infty$). Let us first neglect the boundary 
scattering between the chains, making them completely decoupled. Each chain has one zero-energy ABS located at $x=1$, corresponding to an unpaired, or ``dangling'', MF. 
Next, we turn on the interchain boundary scattering, which will hybridize the MFs, and ask how many of the $N$ MFs will survive.   

The mixing of the dangling MFs can be described phenomenologically by a quadratic Hamiltonian of the form
\begin{equation}
\label{MF-mixing}
  {\hat H}_{mix}=i\sum_{m,n=1}^N T_{mn}\hat{\gamma}_m\hat{\gamma}_n,
\end{equation}
where $\hat{\gamma}_n$ are the Majorana operators, satisfying the anticommutation relations $\{\hat{\gamma}_m,\hat{\gamma}_n\}=2\delta_{mn}$, and $T_{mn}$'s form a real antisymmetric matrix. 
Additionally, the above Hamiltonian has to be invariant under the antiunitary operation (\ref{A-Type-III}). According to Appendix \ref{app: MF transformation}, the transformation rules for the Majorana operators read
\begin{equation}
\label{gamma-transform-A}
  {\cal A}\hat{\gamma}_n{\cal A}^{-1}=\sgn(\eta_n)\hat{\gamma}_n.
\end{equation}
Using this in Eq. (\ref{MF-mixing}), we obtain the following antiunitary symmetry-imposed constraint on the mixing matrix: $T_{mn}=-\sgn(\eta_m\eta_n)T_{mn}$. Therefore an ${\cal A}$-invariant boundary scattering 
can hybridize only MFs from the bands with opposite gap signs, \textit{i.e.}, 
\begin{equation}
\label{T-matrix}
    \hat T=\left(\begin{array}{cc}
          0 & \hat T_{+-} \\
          \hat T_{-+} & 0  \\ 
         \end{array}\right),
\end{equation}
where $\hat T_{ss'}$ is a $N_{s}\times N_{s'}$ matrix and $\hat T_{-+}=-\hat T_{+-}^\top$. Next, we observe that any real antisymmetric matrix can be brought by an orthogonal transformation $\hat{Q}$ 
to a canonical block-diagonal form:
$$
	 \hat{Q}^{-1}\hat T\hat{Q}=\diag\left\{ 0,...,0,\left(\begin{array}{cc}
          0 & \lambda_1 \\
          -\lambda_1 & 0  \\ 
         \end{array}\right),
	 \left(\begin{array}{cc}
          0 & \lambda_2 \\
          -\lambda_2 & 0  \\ 
         \end{array}\right),...\right\},
$$
with real $\lambda$'s, see Ref. \onlinecite{Gantmakher-book}. Substituting this into Eq. (\ref{MF-mixing}), we see that the number of the zero-energy MFs which are not affected by the interchain boundary scattering is equal 
to the number of zero eigenvalues of $\hat T$. 

While a generic antisymmetric matrix has at least one zero eigenvalue only if it is odd-dimensional, in our case there are more possibilities due to special structure of the 
mixing matrix, see Eq. (\ref{T-matrix}). The analysis is similar to the calculation of the number of zero modes of $\hat W(0)$ given earlier in this subsection. 
The zero-mode eigenvectors of $\hat T$ have the form $(\bm{f}^\top_+,\bm{f}^\top_-)$, where $\bm{f}_+$ and $\bm{f}_-$ satisfy the equations $\hat T_{+-}\bm{f}_-=\bm{0}_+$ and $\hat T_{-+}\bm{f}_+=\bm{0}_-$.
If $N_+>N_-$, then $\bm{f}_-=\bm{0}_-$, while the second equation has $N_+-N_-$ nontrivial solutions. 
If $N_+<N_-$, then $\bm{f}_+=\bm{0}_+$, while the first equation has $N_--N_+$ nontrivial solutions. If $N_+=N_-$, then the mixing matrix is even-dimensional and has no zero eigenvalues.
Thus the number of the zero-energy MFs which are immune against the symmetry-preserving boundary scattering is equal to $|N_+-N_-|$, in agreement with Eq. (\ref{N0-Type III}).

The conclusion that a Type III superconducting wire can have any integer number of the zero-energy MFs is also supported by a topological argument. The antiunitary symmetry ${\cal A}={\cal K}\sigma_y$ 
squares to $+1$, placing a Type III superconductor into the symmetry class BDI, which is characterized in 1D by a $\mathbb{Z}$ invariant.\cite{tenfold-way,FHAB11} Our expression (\ref{N0-Type III}) gives an 
explicit form of this invariant. We would like to note that 
the possibility of a $\mathbb{Z}$ topological invariant in certain models of superconducting nanowires has been previously argued in Refs. \onlinecite{TS12} and \onlinecite{ST13}.
The $\mathbb{Z}_2$ invariant proposed in Ref. \onlinecite{Kit01}, see also Refs. \onlinecite{PL10} and \onlinecite{LSDS11}, is more ``coarse-grained'' in the sense that it does not take into account the fact that
the antiunitary symmetry forbids certain couplings between the MFs, see Eq. (\ref{T-matrix}).

\section{Summary}
\label{sec: summary}

We have developed a theory of normal-state and superconducting properties of quasi-1D quantim wires, in which the full 3D inversion symmetry is broken by a substrate. The effects of the electron-lattice SO coupling and TR symmetry 
breaking have been studied in a model-independent way, by using general symmetry arguments. 

The symmetry of a quasi-1D noncentrosymmetric crystal is described by one of the sixteen magnetic point groups, or magnetic classes, with only ten of them consistent with a uniform magnetic field (or a uniform magnetization). 
Superconductivity of the BCS type can potentially exist in eleven magnetic classes. The magnetic classes fall into one of three types, depending on the way the TR operation ${\cal K}$ enters the symmetry elements. 
The electronic band structure, in particular,
the number and location of the Fermi points, is qualitatively different for different types of magnetic classes. As a by-product, we have shown how a generalized Rashba model can be derived in the quasi-1D case.  

We have emphasized the crucial role of antiunitary symmetries in ensuring that the superconducting gap function has the right transformation properties. 
The standard approach of superconductivity theory, in which the Cooper pairs are built from time-reversed states, see Refs. \onlinecite{pseudospin} and \onlinecite{Blount85}, 
has been extended to the case of an arbitrary magnetic symmetry. We have identified the appropriate antiunitary symmetry operations ${\cal A}$ for Type II and Type III classes. 
Regardless of the pairing mechanism, superconductivity in Type II systems corresponds to a quasi-1D multiband version of a conventional $s$-wave pairing.
In contrast, Type III systems exhibit a ``$p$-wave'' pairing and thus represent multiband generalizations of the Kitaev model.

We have studied the spectrum of the Andreev bound states near the end of a superconducting wire using the semiclassical approach. The boundary conditions for the Andreev equations are formulated phenomenologically, 
in terms of the normal-state boundary scattering matrix. In a superconducting state invariant under the antiunitary symmetry ${\cal A}$, we have calculated the number of the
zero-energy ABS which are protected against symmetry-preserving perturbations. In the Type II case, we have ${\cal A}={\cal K}$, therefore ${\cal A}^2=-1$ (class DIII, according to the ``tenfold'' classification of 
Ref. \onlinecite{tenfold-way}) and the number of zero modes is given by 
the $\mathbb{Z}_2$ invariant (\ref{N0-Type II}). In the Type III case, we have ${\cal A}={\cal K}\sigma_y$, therefore ${\cal A}^2=+1$ (class BDI), and the number of zero modes is given by the $\mathbb{Z}$ invariant 
(\ref{N0-Type III}).

Our work might be extended in several directions. One obvious omission is Type I case, in which the lack of an antiunitary symmetry in the magnetic class makes analysis more complicated. In particular,
the classification of the gap functions according to their parity is no longer possible. Also, our counting of the zero-energy ABS is expected to be significantly modified by the interactions beyond the mean-field treatment of 
the pairing. Such interactions would lead to higher-order mixing terms in the MF Hamiltonian, thus reducing the number of protected zero modes.\cite{top-class-interactions}

\acknowledgments
This work was supported by a Discovery Grant from the Natural Sciences and Engineering Research Council of Canada.

\appendix

\section{Spin representation of pairing}
\label{app: gap spin}

Suppose we have a Type II or Type III quantum wire with two single-electron bands described by the Rashba model. Using the relations
$$
  \hat c^\dagger_{k_x\lambda}=\sum_{\alpha=\uparrow,\downarrow}\hat a^\dagger_{k_x\alpha}\langle k_x\alpha|k_x\lambda\rangle,\quad
  \hat{\tilde c}^\dagger_{k_x\lambda}=\sum_{\alpha=\uparrow,\downarrow}\hat{\tilde a}^\dagger_{k_x\alpha}\langle k_x,\lambda|k_x,\alpha\rangle,
$$
where $\lambda=\pm$ and $\hat{\tilde a}^\dagger_{k_x\alpha}={\cal A}\hat a^\dagger_{k_x\alpha}{\cal A}^{-1}=\sum_\beta\hat a^\dagger_{-k_x,\alpha}A_{\beta\alpha}$, cf. Eq. (\ref{A-matrix-band}), 
the pairing Hamiltonian can be written as 
$$
  \sum_{k_x,\lambda}\tilde\Delta_\lambda(k_x)\hat c^\dagger_{k_x\lambda}\hat{\tilde c}^\dagger_{k_x\lambda}=\sum_{k_x,\alpha\beta}\Delta_{\alpha\beta}(k_x)\hat a^\dagger_{k_x\alpha}\hat a^\dagger_{-k_x,\beta}. 
$$
The gap function on the right-hand side is a $2\times 2$ spin matrix: $\hat{\Delta}(k_x)=\sum_{\lambda}\tilde\Delta_\lambda(k_x)\hat\Pi_{\lambda}(k_x)\hat A^\top$, where 
$$
  \hat\Pi_\lambda(k_x)=|k_x,\lambda\rangle\langle k_x,\lambda|=\frac{1+\lambda\hat{\bGam}(k_x)\hat{\bm{\sigma}}}{2}
$$
is the projection operator onto the $\lambda$th helicity band, with $\hat{\bGam}=\bGam/|\bGam|$, see Eq. (\ref{Gamma-def}). Therefore
\begin{eqnarray}
\label{gap-matrix}
  \hat{\Delta}(k_x) &=& \frac{\tilde\Delta_+(k_x)+\tilde\Delta_-(k_x)}{2}\hat A^\top+\frac{\tilde\Delta_+(k_x)-\tilde\Delta_-(k_x)}{2}\hat{\bGam}(k_x)(\hat{\bm{\sigma}}\hat A^\top)\nonumber\\
	&=& d_0(k_x)(i\hat\sigma_2)+\bm{d}(k_x)(i\hat{\bm{\sigma}}\hat\sigma_2).
\end{eqnarray}
In the second line we used the standard representation of the gap function in terms of the spin-singlet component $d_0(k_x)=d_0(-k_x)$ and the spin-triplet components $\bm{d}(k_x)=-\bm{d}(-k_x)$, 
see, \textit{e.g.}, Ref. \onlinecite{Book}.

For Type II classes, it follows from Eq. (\ref{A-Type-II}) that $\hat{\tilde a}^\dagger_{k_x\alpha}=i\sigma_{2,\alpha\beta}\hat a^\dagger_{-k_x,\beta}$, therefore $\hat A=-i\hat\sigma_2$. Since in this
case $\bGam(k_x)=\bgam(k_x)$, we obtain from Eq. (\ref{gap-matrix}):
\begin{equation}
\label{psi-d-II}
  d_0(k_x)=\frac{\tilde\Delta_+(k_x)+\tilde\Delta_-(k_x)}{2},\quad \bm{d}(k_x)=\frac{\tilde\Delta_+(k_x)-\tilde\Delta_-(k_x)}{2}\hat{\bgam}(k_x),
\end{equation}
where $\tilde\Delta_\pm$ are even in $k_x$, according to Eq. (\ref{Delta-Type-II-parity}). The second of these expressions describes the ``protected'' triplet pairing, which survives the large SO band splitting.\cite{NCSC-book}

For Type III classes, it follows from Eq. (\ref{A-Type-III}) that $\hat{\tilde a}^\dagger_{k_x\alpha}=\hat a^\dagger_{-k_x,\alpha}$, therefore $\hat A=\hat\sigma_0$. The gap function (\ref{gap-matrix}) 
takes the form
\begin{equation}
\label{Delta-III-Rashba}
  \hat\Delta(k_x)=\frac{\tilde\Delta_+(k_x)+\tilde\Delta_-(k_x)}{2}+\frac{\tilde\Delta_+(k_x)-\tilde\Delta_-(k_x)}{2}\frac{\bgam(k_x)\hat{\bm{\sigma}}+\bm{h}\hat{\bm{\sigma}}}{\sqrt{\bgam^2(k_x)+\bm{h}^2}},
\end{equation}
where $\tilde\Delta_\pm$ are odd in $k_x$, according to Eq. (\ref{Delta-Type-III-parity}). Note that $\bgam\parallel\hat{\bm{y}}$ and  $\bm{h}\perp\hat{\bm{y}}$, see Table \ref{table: Rashba-vectors}. 
In the limit of strong Zeeman field, $|\bm{h}|\gg|\bgam|$, the contribution of the asymmetric SOC to the band splitting can be neglected and we obtain:
$$
  \hat\Delta(k_x)=\frac{\tilde\Delta_+(k_x)+\tilde\Delta_-(k_x)}{2}+\frac{\tilde\Delta_+(k_x)-\tilde\Delta_-(k_x)}{2}\frac{\bm{h}\hat{\bm{\sigma}}}{|\bm{h}|}.
$$
This expression has an entirely expected form, since in the large-$\bm{h}$ limit the difference between the helicity representation and the spin representation disappears. The electron spins in the two nondegenerate bands 
are fully polarized either parallel or antiparallel to $\bm{h}$, and the same-spin pairing is described by $\tilde\Delta_+(k_x)$ or $\tilde\Delta_-(k_x)$, respectively.

\section{Antiunitary symmetry of the $S$ matrix}
\label{app: S-matrix-symmetry}

In order to obtain the constraints imposed on the $S$ matrix by the antiunitary symmetry of the normal state, 
we express the wave function of normal electrons away from the boundary as a superposition of $N$ incident and $N$ reflected states:
\begin{equation}
\label{Psi-general-N}
  |\Psi\rangle=\sum_{n=1}^N\left(C_n|k^{\mathrm{in}}_n\rangle+\tilde C_n|k^{\mathrm{out}}_n\rangle\right),
\end{equation}
where $|k\rangle\equiv|k_x,n\rangle$ is a shorthand notation for the Bloch spinor state corresponding to the Fermi wavevector $k_x=k^{\mathrm{in}}_n$ or $k^{\mathrm{out}}_n$. 
The scattering matrix is defined by the following relation between the coefficients:  
\begin{equation}
\label{S-matrix-N}
  \tilde C_n=\sum_m S_{nm}C_m.
\end{equation}
Due to the particle number conservation in the boundary scattering, the $S$ matrix is unitary.

Applying the antiunitary operation ${\cal A}$ to the wave function (\ref{Psi-general-N}) and using Eqs. (\ref{A-matrix-band}) and (\ref{t_n-def}), we obtain: 
\begin{eqnarray*}
  {\cal A}|\Psi\rangle &=& \sum_n \left[C_n^*t_n(k^{\mathrm{in}}_n)|-k^{\mathrm{in}}_n\rangle+\tilde C_n^*t_n(k^{\mathrm{out}}_n)|-k^{\mathrm{out}}_n\rangle\right]\\
  &=& \sum_n \left[\tilde C_n^*t_n(k^{\mathrm{out}}_n)|k^{\mathrm{in}}_n\rangle+C_n^*t_n(k^{\mathrm{in}}_n)|k^{\mathrm{out}}_n\rangle\right].
\end{eqnarray*}
Note how the ``in''- and ``out''-states are interchanged by ${\cal A}$. Since the normal-state Hamiltonian, including the boundary, is invariant under ${\cal A}$, one can expect the same 
$S$-matrix relations between the incident and reflected states in $|\Psi\rangle$ and ${\cal A}|\Psi\rangle$, \textit{i.e.},
\begin{equation}
\label{TR-S-matrix-N}
  C_n^*t_n(k^{\mathrm{in}}_n)=\sum_m S_{nm}\tilde C_m^*t_m(k^{\mathrm{out}}_{m}).
\end{equation}
Comparing Eqs. (\ref{S-matrix-N}) and (\ref{TR-S-matrix-N}) and taking into account the unitarity of the $S$-matrix, we obtain:
$$
  S_{mn}=t_n^*(k^{\mathrm{in}}_n)S_{nm}t_m(k^{\mathrm{out}}_{m})=t_n^*(-k^{\mathrm{out}}_n)S_{nm}t_m(k^{\mathrm{out}}_{m}).
$$
Depending on whether ${\cal A}$ squares to $-1$ or $+1$, the phase factor $t_n$ is an odd or even function of $k_x$, see Secs. \ref{sec: Type II-gap} and \ref{sec: Type III-gap}. Therefore
\begin{equation}
\label{TR-S}
  S_{mn}=\mp t_n^*(k^{\mathrm{out}}_n)S_{nm}t_m(k^{\mathrm{out}}_{m}),
\end{equation}
where the upper (lower) sign corresponds to the Type II (Type III) case.

The phase factors $t_n$ depend on the phase choice for the Bloch states. In particular, one can make $t_n(k_x)$ real and equal to $+1$ for $k_x=k^{\mathrm{out}}_{n}$. 
Then we obtain from Eq. (\ref{TR-S}) that in this basis the scattering matrix is either antisymmetric:
\begin{equation}
\label{S-Type-II}
  \hat S^\top=-\hat S\qquad (\mathrm{Type\ II}),
\end{equation}
or symmetric: 
\begin{equation}
\label{S-Type-III}
  \hat S^\top=\hat S\qquad (\mathrm{Type\ III}).
\end{equation}
It follows from Eq. (\ref{S-Type-II}) that $S_{nn}=0$, \textit{i.e.}, the backscattering of quasiparticles into the same band is forbidden by TR symmetry.

\section{Antiunitary symmetry of Majorana fermions}
\label{app: MF transformation}

Consider a single Kitaev chain described by the Hamiltonian
\begin{equation}
\label{Kitaev-chain}
  \hat H=\sum_x\left(-t\hat c_x^\dagger \hat c_{x+1}-\eta\hat c_x\hat c_{x+1}+\mathrm{H.c.}\right)-\mu\sum_x\hat c_x^\dagger \hat c_x,
\end{equation}
where $x=1,...,L$ labels the 1D lattice sites, $t$ and $\mu$ are, respectively, the nearest-neighbor hopping and the chemical potential, and $\eta=|\eta|e^{i\theta}$ is the superconducting order parameter. 
The momentum representation of Eq. (\ref{Kitaev-chain}) reproduces the gap function (\ref{Type-III-OP}), with $\phi(k_x)=\sin(k_xd)$, where $d$ is the lattice spacing.
Following Ref. \onlinecite{Kit01}, we introduce $2L$ Majorana operators
\begin{equation}
\label{MF-definition}
    \hat\tau_{2x-1}=e^{i\theta/2}\hat c_x+e^{-i\theta/2}\hat c_x^\dagger,\quad 
    \hat\tau_{2x}=-ie^{i\theta/2}\hat c_x+ie^{-i\theta/2}\hat c_x^\dagger.
\end{equation}
In the nontrivial phase, with $|\eta|\neq 0$ and $|\mu|<2|t|$, there is one unpaired MF per each end of the chain, which are described by the operators $\hat\gamma=\hat\tau_1$ and $\hat\gamma'=\hat\tau_{2L}$. 

Since the action of the antiunitary operation (\ref{A-Type-III}) is given by ${\cal A}(f\hat c_x){\cal A}^{-1}=f^*\hat c_x$ and ${\cal A}(f\hat c_x^\dagger){\cal A}^{-1}=f^*\hat c_x^\dagger$, where 
$f$ is a $c$-number, an ${\cal A}$-invariant 
superconducting state corresponds to a real $\eta$. If $\eta$ is real positive ($\theta=0$), then one obtains from Eq. (\ref{MF-definition}):
$$
   {\cal A}\hat\tau_{2x-1}{\cal A}^{-1}=\hat\tau_{2x-1},\quad {\cal A}\hat\tau_{2x}{\cal A}^{-1}=-\hat\tau_{2x},  
$$
but if $\eta$ is real negative ($\theta=\pi$), then 
$$
   {\cal A}\hat\tau_{2x-1}{\cal A}^{-1}=-\hat\tau_{2x-1},\quad {\cal A}\hat\tau_{2x}{\cal A}^{-1}=\hat\tau_{2x}. 
$$
Focusing on the unpaired MF at $x=1$, we have 
\begin{equation}
\label{MF-end-transform}
  {\cal A}\hat\gamma{\cal A}^{-1}=\sgn(\eta)\,\hat\gamma.
\end{equation}
Therefore the response of the MF on the antiunitary symmetry operation depends on the sign of the order parameter. This becomes important if there is more than one Kitaev chain, with different signs of the gap.
Introducing the chain index $n$ in Eq. (\ref{MF-end-transform}), we arrive at Eq. (\ref{gamma-transform-A}).

\clearpage

\clearpage

\clearpage

\end{document}